\documentclass[preprint]{aastex62}

\graphicspath{{./}{figures/}}

\usepackage{graphicx}
\usepackage[caption=false]{subfig}

\newcommand{\chandra}{{\it Chandra}}

\newcommand{\OIII}{[O\,{\tiny III}]}

\begin{document}

\title{A {\it Chandra} X-ray Survey of Optically Selected AGN Pairs}

\author[0000-0001-9062-8309]{Meicun Hou}
\affiliation{School of Astronomy and Space Science, Nanjing University, Nanjing 210046, China}
\affiliation{Key Laboratory of Modern Astronomy and Astrophysics (Nanjing University), Ministry of Education, Nanjing 210046, China}
\affiliation{Department of Astronomy, University of Illinois at Urbana-Champaign, Urbana, IL 61801, USA}
\email{houmc@smail.nju.edu.cn}

\author[0000-0003-0355-6437]{Zhiyuan Li}
\affiliation{School of Astronomy and Space Science, Nanjing University, Nanjing 210046, China}
\affiliation{Key Laboratory of Modern Astronomy and Astrophysics (Nanjing University), Ministry of Education, Nanjing 210046, China}
\email{lizy@nju.edu.cn}

\author[0000-0003-0049-5210]{Xin Liu}
\affiliation{Department of Astronomy, University of Illinois at Urbana-Champaign, Urbana, IL 61801, USA}
\affiliation{National Center for Supercomputing Applications, University of Illinois at Urbana-Champaign, 605 East Springfield Avenue, Champaign, IL 61820, USA}
\email{xinliuxl@illinois.edu}

\begin{abstract}
We present a {\it Chandra} archival study of optically selected AGN pairs at a median redshift $\bar{z} \sim$ 0.1. 
Out of 1286 AGN pairs (with projected separations $r_p < 100$ kpc and velocity offsets $\Delta v < 600$ km s$^{-1}$) optically identified from the Sloan Digital Sky Survey Seventh Data Release, we find 67 systems with archival Chandra observations, which represent the largest sample of optically selected AGN pairs studied in the X-ray. Among the 67 AGN pairs, 21 systems have both nuclei detected in the X-rays, 36 have one nucleus detected in the X-rays, whereas 10 have no X-ray detection. The X-ray detection rate, 78/134=58\%($\pm7$\% 1$\sigma$ Poisson errors), is significantly higher than that (23/134=17\%$\pm4$\%) of a comparison sample of star-forming galaxy pairs, lending support to the optical AGN classification. 
In the conservative case where X-ray contamination from star formation is removed, the X-ray detection rate becomes $27\% \pm 4\%$, consistent with predictions from latest galaxy merger simulations.  
The 2--10 keV X-ray luminosity $L_{2-10~{\rm keV}}$ increases with decreasing projected separation in AGN pairs for $r_p \gtrsim 15$ kpc, suggesting an enhancement of black hole accretion even in early-stage mergers. On the other hand, $L_{2-10~{\rm keV}}$ appears to decrease with decreasing projected separation at $r_p \lesssim 15$ kpc, which is contradictory to predictions from merger simulations. The apparent decrease in $L_{2-10~{\rm keV}}$ of AGN pairs at $r_p \lesssim 15$ kpc may be caused by: (i) enhanced absorbing columns from merger-induced gas inflows, (ii) feedback effects from early-stage mergers, and/or (iii) small number statistics. Future X-ray studies with larger samples are needed to put our results on firmer statistical ground.
\end{abstract}


\section{Introduction} \label{sec:intro}

Galaxy mergers and close interactions have long been recognized to have played a vital role in shaping the structure and morphology of galaxies \citep[e.g.,][]{zwicky56,vorontsov59,toomre72,khachikian87,sanders88,barnes92} and particularly the cores of ellipticals \citep[e.g.,][]{faber97,graham04,merritt04,lauer05,Kormendy2009}. They are thought to drive accretion in the central supermassive black holes \citep[SMBHs;][]{hernquist89,norman88,moore96,ciotti07,ciotti10}, fueling at least some Active Galactic Nuclei (AGNs). However, the causal link between galaxy interactions and AGNs has been highly controversial. The details of AGN feeding, and in particular, the role of mergers, may be radically different at high and low accretion rates \citep[e.g.,][]{ho08,Treister2012,Hopkins2014,Villforth2017,Donley2018}.

One fundamental question in the study of AGNs is the mechanism for triggering their growth. For quasars (i.e., high-luminosity AGNs), one hypothesis is that they are triggered by gas-rich major mergers between galaxies \citep[the ``merger hypothesis'';][]{sanders96,kauffmann00,hopkins08}. In the hierarchical $\Lambda$CDM paradigm, small structures merge to form large structures, and the merger rate of dark matter halos peaks at early times \citep{bell06}, in broad consistency with the observed peak of bright quasar activities \citep[e.g.,][]{richards06}. Gas-rich major mergers (i.e., those between galaxies of comparable mass and with plenty of gas) provide an efficient mechanism for channelling a large amount of gas into the central region, triggering starbursts and rapid black hole growth \citep[e.g.,][]{hernquist89,mihos96,dimatteo05}. Observationally, the merger hypothesis is supported by the connection between quasars and ultra-luminous infrared galaxies \citep[ULIRGs; e.g.,][]{sanders96}, by signatures of recent mergers in quasar hosts \citep[e.g.,][]{bahcall97,bennert08}, by the small-scale over-densities of galaxies around luminous quasars and quasar pairs \citep[e.g.,][]{bahcall97,serber06,Hennawi2006,myers08,shen10c}, and by the obscuration in the central region of the merger systems \citep[e.g.,][]{urrutia08,Satyapal2014,Fanlulu2016,Ricci2017}. While these observations do not necessarily prove that mergers are directly responsible for quasar triggering, they suggest that quasar activity is coincident with mergers in many cases. Indeed, merger-based phenomenological models can reproduce virtually all the statistical properties of quasars \citep[e.g.,][]{hopkins08,shen09}.

However, at lower AGN luminosities (i.e., $L_{\rm bol}\lesssim 10^{45}\,{\rm erg\,s^{-1}}$), the causal link between mergers and AGN triggering has been elusive. Some observations found an excess of close neighbors in AGN host galaxies or a higher fraction of AGNs in interacting than in isolated galaxies \citep[e.g.,][]{petrosian82,hutchings83,dahari84,kennicutt84,keel85,bahcall97,serber06,koss10,silverman11,ellison11,Liu2012} whereas others detected no significant difference \citep[e.g.,][]{dahari85,schmitt01,miller03,grogin05,waskett05,pierce07,ellison08,Li2008,darg10a,Villforth2014,Villforth2017}. Because less fuel is required, secular processes such as instabilities driven by bars and minor mergers, may be sufficient to trigger low-luminosity AGNs \citep[e.g.,][]{Hopkins2014,Menci2014}. One important goal in AGN studies is to distinguish between different fueling routes, i.e., violent major-mergers versus secular processes, and to assess their relative contributions as functions of redshift and AGN luminosity.

Binary SMBHs are thought to be a generic outcome in the hierarchical paradigm of structure formation \citep{begelman80,milosavljevic01,yu02}, given that most massive galaxies harbor SMBHs \citep{kormendy95,richstone98}. Binary SMBHs have long been proposed theoretically \citep[e.g.,][]{begelman80,milosavljevic01,yu02,DEGN}, invoked to explain a variety of phenomena, such as the core properties of elliptical galaxies \citep[e.g.,][]{faber97,Kormendy2009}, ``X''-shaped radio morphologies \citep[e.g.,][]{merritt02b}, peculiar AGN broad emission line profiles \citep[e.g.,][]{gaskell82,boroson09,shen10,eracleous11,shen13,Liu2014,Wang2017}, quasi-periodic flux variabilities in AGNs \citep[e.g.,][]{Graham2015,Graham2015a,DOrazio2015a}. Both SMBHs in a merger may accrete at the same time, which will be observable as a pair of AGNs (also being referred to as ``dual AGNs'' for pairs with separations $<$10 kpc; for a recent review, see \citealt{DeRosa2020}). However, it is difficult to predict when one and in particular both SMBHs become active \citep[e.g.,][]{shlosman90,Armitage2002,wada04,Dotti2007,Dotti2012,Blecha2013a}. Observations of AGN pairs help understand galaxy mergers in general and tidally-triggered AGN in particular. Their host-galaxy properties offer clues to the external and internal conditions under which both SMBHs are activated at the same time.

Large, homogeneous optical surveys such as the Sloan Digital Sky Survey (SDSS) have increased the inventory of AGN pairs on tens-of-kpc to kpc scales by more than an order of magnitude. At high redshift ($z>0.5$), quasar pair candidates can be selected from imaging data and confirmed by follow-up spectroscopy. Due to the moderate imaging resolution and the relatively high redshift, most SDSS quasar pairs are on $>$ tens-of-kpc scales \citep[e.g., tens of systems known with $\sim$50--100 kpc (projected separation) at $0.5<z<5$;][]{Hennawi2006,Hennawi2010,shen10c,Vignali2018}. However, these quasar pairs are still relatively far away from each other; it is unclear that their host galaxies are in direct tidal interactions with each other or not, except in rare cases where tidal tails are seen using deep imaging at the low-redshift end \citep[e.g.,][]{green10}. At lower redshift ($z<0.5$), AGN pairs can be found at smaller physical separations where direct tidal interactions between their host galaxies are more likely. 

However, studies based on large ($>10^3$ pairs), optical AGN pair samples are hampered by the use of \OIII\ luminosity as an AGN luminosity surrogate. \OIII\ comes from the narrow-line regions which are thought to reside outside the dusty torus \citep{antonucci93}; there is a large scatter ($\sim$1 dex) between \OIII\ luminosity and the intrinsic AGN luminosity \citep[e.g.,][]{heckman05,panessa06,Lamastra2009}. In contrast, X-rays come from the corona much closer to the SMBH and accretion disk \citep{Haardt1991,Haardt1993}, providing arguably the most direct evidence for AGN. In particular, 
the hard X-ray (2--10 keV) emission can still be detectable even with the obscuring torus (column densities $N_{{\rm H}}\lesssim10^{24}$) and is powerful to reveal obscured AGNs except in Compton-thick sources \citep{Pier1992}. {\it Chandra} was instrumental in discovering the prototypical dual AGN in the ULIRG NGC 6240 \citep{komossa03,Tananbaum2014}, whose nuclei are heavily obscured in the optical and are characteristic of LINERs \citep{lutz99}; another example is the dual AGN in Mrk 266 \citep{mazzarella11}. Similarly, many nuclei in the optically selected AGN pair sample of \citet{Liu2011a} have line ratios characteristic of LINERs or composites, which may be driven by stellar or shock heating \citep[e.g.,][]{lutz99,terashima00,Yan2012}. X-rays provide the most unambiguous tracer of AGNs. Previous X-ray studies, however, are limited to relatively small samples of local ULIRGs, Swift BAT AGNs or C-GOALS \citep[e.g.,][]{teng12,koss12,Koss2016,Iwasawa2011,Torres-Alba2018} which represent the low-luminosity AGN population in nearby galaxies. The main challenge in X-ray studies is the need for a statistically large sample that covers a wide dynamical range in spatial separations.  

In this paper, we present an X-ray study of a parent sample of 1286 \OIII -selected (including both type 1s, i.e., unobscured and type 2s, i.e., obscured) AGN pairs at redshift $z\sim0.1$ matched against the \chandra\ X-ray data archive. Our targets were optically selected from the SDSS DR7 \citep{SDSSDR7} by \citet{Liu2011a} which represent the largest sample of optically selected AGN pairs. Both the nuclei of each of these galaxy mergers are optically classified as type 2 or type 1 AGNs\footnote{While the parent sample includes type 1, the parent sample is largely dominated by type 2 AGNs.}. The optical classification for the two nuclei is based on their individual optical spectra from the SDSS, which exhibit optical diagnostic emission line ratios indicative of AGN (including both Seyferts and LINERs) and/or AGN-H\,{\tiny II} composites. We use the X-ray data from {\it Chandra} to examine the nature of the ionizing sources in each system and measure the intrinsic X-ray luminosity for each nucleus. Our results provide the most statistically significant evidence for the elusive link between mergers and AGN in the X-rays. A large sample and a large dynamic range in merger separation are both crucial for discovering any subtle separation-dependent effects expected in the moderate AGN luminosity regime \citep[e.g.,][]{Liu2012}.


In addition, we present an X-ray census of (i) a parent sample of $\sim10^4$ SDSS star-forming galaxies with close companions and (ii) a parent sample of $\sim10^5$ SDSS single AGNs in isolated galaxies. These will be compared with the X-ray properties of optically selected AGN pair sample to put our results into context. 

The paper is organized as follows. Section~\ref{sec:data} describes our sample selection of the optical selected AGN pairs with archival X-ray observations and the construction of the comparison sample of star-forming galaxy pairs and the control sample of single AGNs in isolated galaxies. We present our method and analysis in Section~\ref{sec:analysis} and the results in Section~\ref{sec:result}. Finally, we discuss the implications of our results in Section~\ref{sec:discussion} and summarize in Section~\ref{sec:summary}. 
Throughout this work, we quote errors at 68\% confidence level unless otherwise noted.

\section{Sample Selection}\label{sec:data}


\subsection{Optically Selected AGN Pairs with Archival Chandra Observations}

We start from the 1286 optically selected AGN pairs or multiples at a median redshift $\bar{z} \sim 0.1$ from \citet{Liu2011a}. It was selected with projected separations $r_p < 100$ kpc and line-of-sight velocity offsets $\Delta v < 600$ km s$^{-1}$ from a parent sample of 138,070 AGNs spectroscopically identified based on their optical diagnostic emission-line ratios \citep[i.e., the BPT diagram][]{bpt,veilleux87,kewley01} and widths from the SDSS DR7 \citep[]{SDSSDR7}. We cross-matched this sample with the {\it Chandra} X-ray data archive for observations taken with the Advanced CCD Imaging Spectrometer (ACIS) which are publicly available as of May 2019. We requested both AGNs in a pair fall within $8\arcmin$ from the aimpoint and are covered by the ACIS field-of-view (FoV) to ensure both the observability and optimal angular resolution. We then visually checked all the matched targets, further excluding AGN pairs that fall on the ACIS-S4 CCD and a confirmed triple AGN (SDSS J0849+1114) recently discovered by \citealt[][see also \citealt{Pfeifle2019a}]{Liu2019}. 

Our final sample consists of 67 AGN pairs with {\it Chandra} archival observations from the optically selected AGN pairs. Among them, 22 pairs are observed as the prime target. These include one triple AGN candidate SDSS J0858+1822 observed in {\it Chandra} Program GO-14700279 (PI: Liu), one triple AGN candidate SDSS J1027+1749 optically identified by \citet{Liu2011}\footnote{The third nucleus of these two candidate triple AGN are not included in SDSS DR7, so we only count the separation between nuclei A and B.} and five systems of AGN pairs studied by \citet{Hou2019}. The remaining 45 pairs were serendipitously observed. 
We address potential selection bias and incompleteness in Section~\ref{sec:discussion}.



\subsection{Comparison and Control Samples}
\label{subsec:control}

To put into context the X-ray detection rate of optically selected single/dual AGNs,
we construct a comparison sample of star-forming galaxy (SFG) pairs from the SDSS legacy survey. These galaxy pairs are identified with the same criteria of projected separations $r_p <100$ kpc and velocity offsets $<600$ km s$^{-1}$, but are spectroscopically classified as H\,{\tiny II} nuclei using the BPT diagram. We have visually inspected all the pairs with small separations ($<$5--10 kpc) to remove interlopers due to multiple fiber positions within the same galaxy with $\sim$3000 SFG pairs left. We then cross-matched this parent list with the {\it Chandra} archive using the same criteria as for the AGN pairs. This resulted in a sample of 67 SFG pairs (the number is just coincidentally identical with the number of AGN pairs).
There are only 2 SFG pairs with $r_p <8$ kpc, in contrast to 15 AGN pairs with $r_p <8$ kpc, which may be due in part to a selection bias (i.e., most close AGN pairs were covered by targeted {\it Chandra} observations). 

To study the effects of merger-induced accretion, we also draw a control sample of 115 single AGNs in isolated galaxies with {\it Chandra} observations that are matched to each nucleus of the AGN pairs in redshift (${\Delta}z \leq 0.004$) and stellar mass (log$M_* \leq 0.05$). The isolated galaxies are defined as galaxies without any companion within a projected distance of 100 kpc. In cases where more than one matches are available for a given nucleus, we randomly pick one of them. The rest, 134$-$115=19 nuclei in AGN pairs do not have matched single AGNs, either because their stellar mass is undetermined, or because the otherwise matched single AGNs have not been observed by {\it Chandra}. 

\begin{figure}\centering
\includegraphics[width=0.35\textwidth,angle=90]{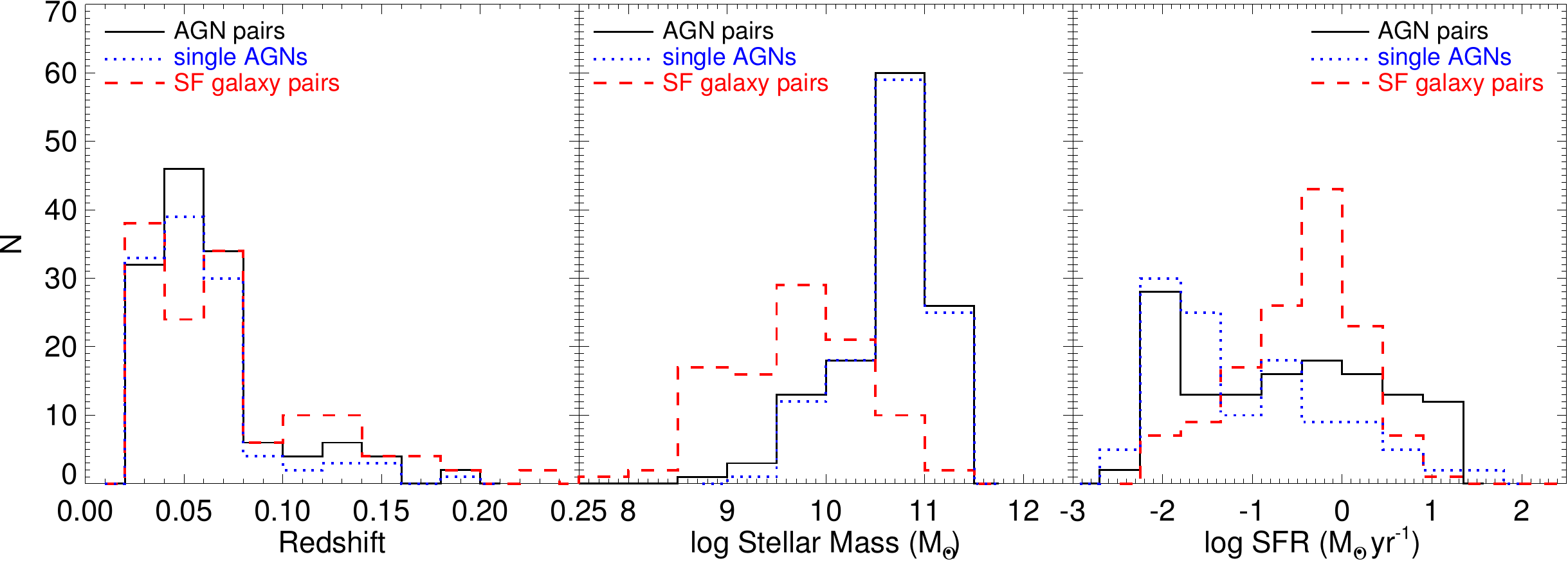}
\caption{Redshift, stellar mass and star formation rate distributions of the AGN pairs and control samples of single AGNs and star-forming galaxy pairs, which are shown in black solid, red dashed and blue dotted histogram respectively.
}
\label{fig:samples}
\end{figure}

Figure \ref{fig:samples} shows the redshift, stellar mass  and star formation rate (SFR) distributions of the AGN pairs and those of the comparison and control samples. 
 We use the SDSS fiber SFR given by the MPA-JHU DR7 catalog \citep{salim07}, which refers to the amount of SFR within the nucleus and is determined by constructing the likelihood distribution of the specific SFR as a function of the 4000 \AA\ break D$_n$(4000) based on the star-forming sample \citep{brinchmann04} multiplied by the stellar mass.
The AGN pairs have a median redshift $\bar{z} \sim 0.05$, a median stellar mass $M_* \sim 4.9\times10^{10}{\rm~M_\odot}$  and a median SFR $\sim 0.28\rm~{M_{\odot}~{yr}^{-1}}$. The control sample of single AGNs in isolated galaxies has similar redshift and stellar mass distributions, which ensures that any difference observed between the AGN pair and the single AGN samples will not be subject to redshift and/or mass-dependent biases. 
The SFR distribution of single AGNs is systematically lower than AGN pairs, which is consistent with the enhancement of star forming activities in AGN pairs.
On the other hand, the comparison sample of SFG pairs, while also having a similar redshift distribution as the AGN pairs, has a systematically smaller stellar mass distribution. This may be partially due to an intrinsic difference between SFGs and the host galaxies of AGNs, because typical AGN host galaxies tend to be more massive than typical SFGs \citep[e.g.,][]{kauffmann03}.
The SFR distribution of SFG pairs is systematically higher than AGN pairs and more concentrated to a value of $\sim 0.4\rm~{M_{\odot}~{yr}^{-1}}$.

\section{Method and Analysis}\label{sec:analysis}

\subsection{\chandra\ Data Reduction
}\label{subsec:X-ray data}

We reprocessed the archival {\it Chandra} data for all samples using CIAO v4.8 with the calibration files of CALDB v4.7.0 following standard procedures\footnote{\url{http://cxc.harvard.edu/ciao/}}. 
Among the whole sample, 54 AGN pairs have only one observation.
For the other 13 AGN pairs with more than one  observations that match our selection criteria, we checked all the available observations. We combined all observations of 6 AGN pairs (SDSS J0320+4123, SDSS J0805+2410, SDSS J0924+0225, SDSS J1259+2757, SDSS J1415+5204, SDSS J1605+1746) to achieve the optimal signal-to-noise ratio (S/N). For the remaining 7 AGN pairs, we only utilized the observation with the longest exposure; the extra observation(s) had either a very short exposure ($<$ 3 ks) or a large off-axis angle, which does not help with the S/N.
We examined the light curves of each observation and filtered time intervals of significant particle background. 
The effective exposure time of each target ranged from 1.8 ks to 592.2 ks. For each observation, we produced counts, exposure, and point-spread-function (PSF) maps on the original pixel scale ($0 \farcs 492 {\rm~pixel^{-1}}$) in the 0.5--2 ($S$), 2--8 ($H$), and 0.5--8 ($F$) keV bands. We have weighted the exposure maps by a fiducial incident spectrum, which is an absorbed power-law with a photon-index of 1.7 and an intrinsic absorption column density $N_{\rm H}$ = $10^{22} {\rm~cm^{-2}}$ for the $H$ band, and an intrinsic absorption column density $N_{\rm H}$ = $10^{21} {\rm~cm^{-2}}$ for the $S$ band. 
For AGN pairs with multiple observations, we calibrated their relative astrometry by matching the centroid of commonly detected point sources with the CIAO tool {\it reproject$\_$aspect}. The counts, exposure and PSF maps of individual observations were then reprojected to a common tangential point, i.e., the optical position of one nucleus in pair, to produce combined images of optimal sensitivity for source detection. 
To ensure optimal sensitivity and accurate positioning for source detection, we have only used the data from the I0, I1, I2 and I3 CCDs for the ACIS-I observations and the S2 and S3 CCDs for the ACIS-S observations.

Table \ref{tab:info} lists the basic properties of AGN pairs. These include the SDSS name, celestial coordinates, redshift, projected separations $r_p$, stellar mass, total \OIII\ flux (uncorrected for contribution from star formation), star formation rate estimates taken from the MPA-JHU DR7 catalog\footnote{\url{https://wwwmpa.mpa-garching.mpg.de/SDSS/DR7/}}, the {\it Chandra} observation ID
and the effective X-ray exposure time.

\begin{figure}\centering
\includegraphics[width=0.45\textwidth,angle=90]{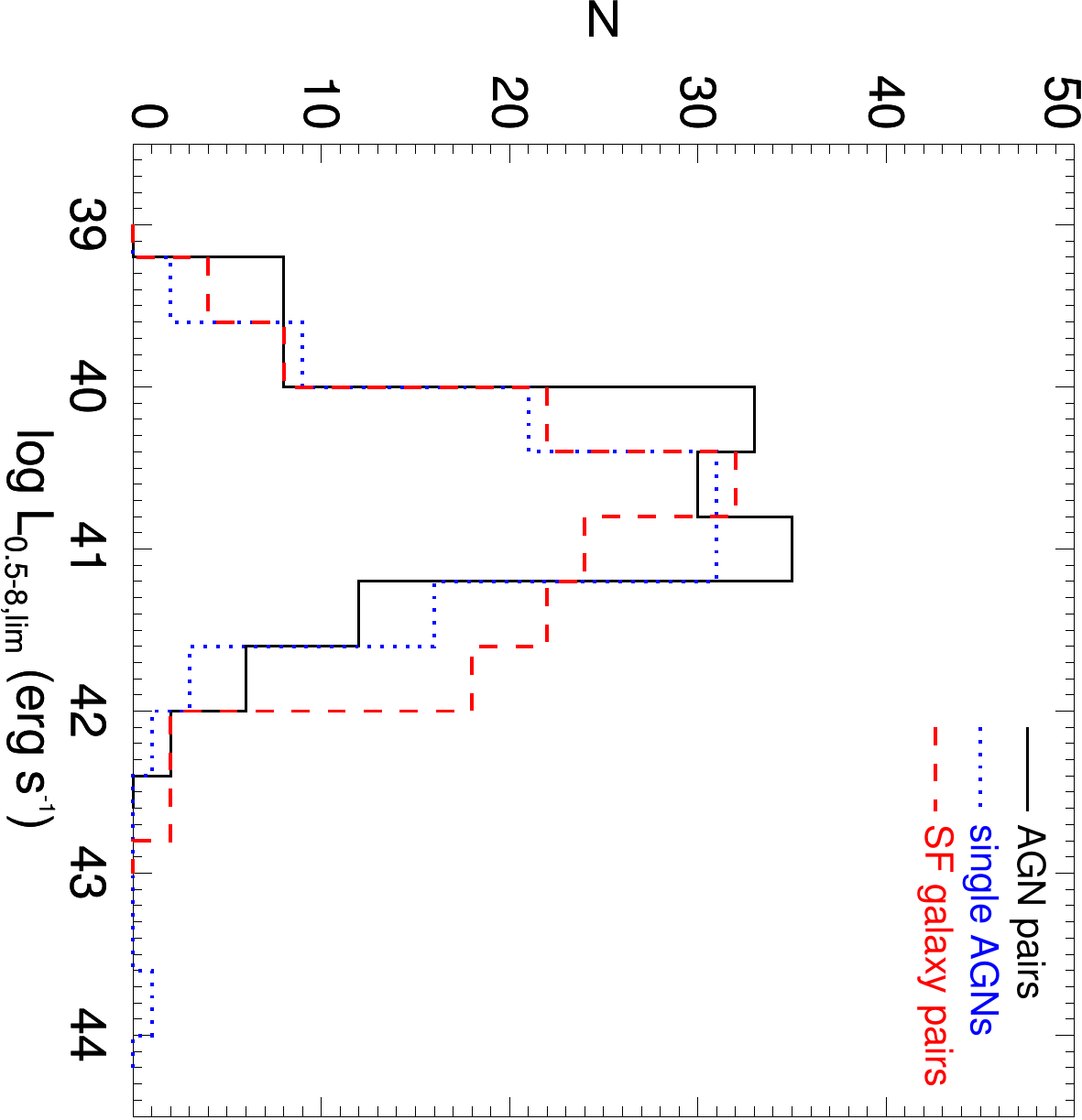}
\caption{Distribution of the 0.5--8 keV ($F$ band) luminosity detection limit of AGN pairs (black solid). Also shown are the control sample of single AGNs in isolated galaxies (blue dotted) and the comparison sample of star-forming galaxy pairs (red dashed).
}
\label{fig:detlim}
\end{figure}

\subsection{X-ray Source Detection}\label{subsec:detection}

We performed X-ray source detection in the 0.5--2, 2--8, and 0.5--8 keV bands for each observation using the CIAO tool {\it wavdetect}. We follow the method described in \citet{Hou2017} and \citet{Jin2019}. We have supplied the exposure and 50\% enclosed count fraction (ECF) PSF maps and have adopted a false detection probability of $P \leq {10^{-6}}$. We have detected a total of 2506 independent X-ray sources in the $F$ band, 1919 in the $S$ band, and 1282 in the $H$ band across the full field of view (FOV). We have also produced the sensitivity maps, which recorded the detection limit at each pixel \citep{Kashyap2010}. We have derived the detection limit at the position of each nucleus from the sensitivity maps.

Table \ref{tab:info} lists the $F$ band detection limit of each target. It ranges between 2.3$\times 10^{-7}$ and $8.7\times 10^{-6} \rm~ph~s^{-1}~cm^{-2}$. 
Figure \ref{fig:detlim} shows the distributions of the detection limit for the three samples, which have been converted to the 0.5--8 keV intrinsic luminosity (see below for details). There is no significant difference in the detection limit among the three samples. 

We use the CIAO tool {\it aprate} to calculate the net photon fluxes within the 90\% enclosed count radius (ECR) for each detected point source in individual bands. In the few cases when two neighboring sources have overlapping 90\% ECR, we adopt a 50\% ECR for photometry. We have subtracted the local background and have corrected for effective exposure and the ECF. The typical background region has been chosen to be between 2 to 5 times the 90\% ECR with any detected X-ray sources removed in the region. 

Assuming the incident power-law spectrum (i.e., with a photon-index of 1.7 and an intrinsic absorption column density $N_{\rm H}$ = $10^{22} {\rm~cm^{-2}}$ for the $H$ band and an intrinsic absorption column density $N_{\rm H}$ = $10^{21} {\rm~cm^{-2}}$ for the $S$ band), we have adopted a photon flux-to-intrinsic energy flux conversion factor of $3.6 \times 10^{-9}{\rm~erg~s^{-1}~cm^{-2}/(ph~s^{-1}~cm^{-2})}$  and $2.2 \times 10^{-9}{\rm~erg~s^{-1}~cm^{-2}/(ph~s^{-1}~cm^{-2})}$  in the 0.5--8 and 0.5--2 keV band and $8.6 \times 10^{-9}{\rm~erg~s^{-1}~cm^{-2}/(ph~s^{-1}~cm^{-2})}$ in the 2--10 keV band. We have applied a K-correction appropriate for sources at different redshifts. To facilitate a direct comparison with previous studies, we have adopted a moderate extrapolation from the 2--8 keV to the 2--10 keV in the luminosity calculation. The assumed canonical absorption column density of $10^{21} {\rm~cm^{-2}}$, while suitable for off-nuclear sources such as X-ray binaries, may be an underestimate for the nuclei if there is substantial intrinsic obscuration. 
In particular, the measured 2--10 keV luminosity can be significantly affected in Compton-thick AGNs, which are most likely found in close or late-stage mergers. But as we will show below, a large fraction of our pairs have a large separation (47 out of 67 pairs with $r_p >$ 10 kpc) and are not in close mergers.
Thus this likely only affects a small fraction of the nuclei in our sample. 

We adopt the same source detection procedures for the {\it Chandra} observations of the comparison and control samples. We have detected a total of 3960 independent sources in the $F$ band, 3223 sources in the $S$ band, and 2023 sources in the $H$ band in the fields of single AGNs in isolated galaxies, and a total of 2147 independent sources in the $F$ band, 1658 sources in the $S$ band, and 1115 sources in the $H$ band in the fields of SFG pairs. The majority of the detected sources may be due to the cosmic X-ray background or the Galactic foreground \citep[e.g.,][]{Hou2017}. A small fraction of the detected sources may also be associated with high-mass X-ray binaries possibly related to merger-induced star formation. We focus our discussion below on the detected nuclear X-ray sources and defer a detailed analysis of the off-nuclear sources to future work. 


\subsection{Identification of X-ray Counterparts}
\label{subsec:identification}

We search for X-ray counterparts of the AGN pairs from the list of detected sources using a matching radius of 2\arcsec . This is chosen in view of the angular resolution of the SDSS imaging and the 90\% ECR of {\it Chandra} in most cases. 
We have performed a random matching test by artificially shifting the positions of all nuclei by $\pm 10\arcsec$ in R.A. and Dec., which results in zero coincident match. As long as an optical nucleus is matched in any of the three energy bands, we consider it detected in the X-rays. For the nuclei that are not automatically picked up by {\it wavdetect}, we use the CIAO tool {\it aprate} to examine the 3-$\sigma$ lower limit of X-ray flux at the optical nucleus position, following the procedures described in Section~\ref{subsec:detection}. If the flux lower limit is greater than zero, the source is regarded as an X-ray detection. From this procedure we have 11 additional nuclear X-ray sources detected in the $F$ band, 10 sources in the $S$ band, and 6 sources in the $H$ band. For the non-detected nuclei, we derive the 3-$\sigma$ flux upper limit using {\it aprate}.
We have thus identified 77 nuclei with X-ray counterparts. 

Visual inspection further identifies one additional nucleus, J122628.28+090126.4, whose SDSS fiber position deviates slightly from the true optical nucleus. A bright X-ray point source is found to be coincident with the nucleus and we include it as an X-ray-detected nucleus. 

In summary, we have identified 78 nuclei with X-ray counterparts in total. Among them, 76 nuclei are detected in the $F$ band, 72 nuclei in the $S$ band, and 49 nuclei in the $H$ band. Table \ref{tab:Xray} lists the net counts in 0.5--8 ($F$) keV band, 0.5--2 keV and 2--8 keV fluxes, and 0.5--2 keV and 2--10 keV intrinsic luminosities for the 78 X-ray-detected nuclei. In the few cases where the X-ray centroids of the nucleus differ significantly in different bands (e.g., the centroid may be problematic because of diffuse X-ray emission in the $S$ band or low count levels in the $H$ band), we take the $F$ band centroid as the fiducial value and refine the photometry for the other bands. We have also identified X-ray nuclei in the control and comparison samples following the same procedure.

\section{Results}\label{sec:result}

\subsection{X-ray Detection Rates}\label{subsec:detectionrate}

Table 3 summarizes the X-ray detection rates. Among the sample of 67 optically selected AGN pairs with X-ray observations, 21 AGN pairs (i.e., $31\%\pm7\%$) have both nuclei detected, 36 AGN pairs have only one of the two nuclei detected, and 10 AGN pairs have no X-ray detection. Figure \ref{fig:image_AGNpair} displays the SDSS optical $gri$ color-composite images and the {\it Chandra} X-ray images of the 21 AGN pairs with both nuclei detected in the X-rays. We find an X-ray detection rate of $58\%\pm7\%$ (78/134) among the 67 AGN pairs. In the subsample of AGN pairs with $r_p \gtrsim 10$ kpc, the X-ray detection rate is $50\%\pm7\%$ (47/94). In the subsample of AGN pairs with $r_p \lesssim 10$ kpc, the X-ray detection rate is significantly larger,  $78\%\pm14\%$ (31/40). In the more conservative case where we only consider X-ray counterparts with $L_{\rm 2-10} > 10^{41} \rm~erg~s^{-1}$ as clearly detected AGNs (see discussion in Section~\ref{subsec:properties}), the X-ray detection rate becomes $27\% \pm 4\%$ (36/134) with 36 nuclei detected in the $H$ band. 

\begin{figure}\centering
\subfloat{\includegraphics[width=0.84\textwidth,angle=0]{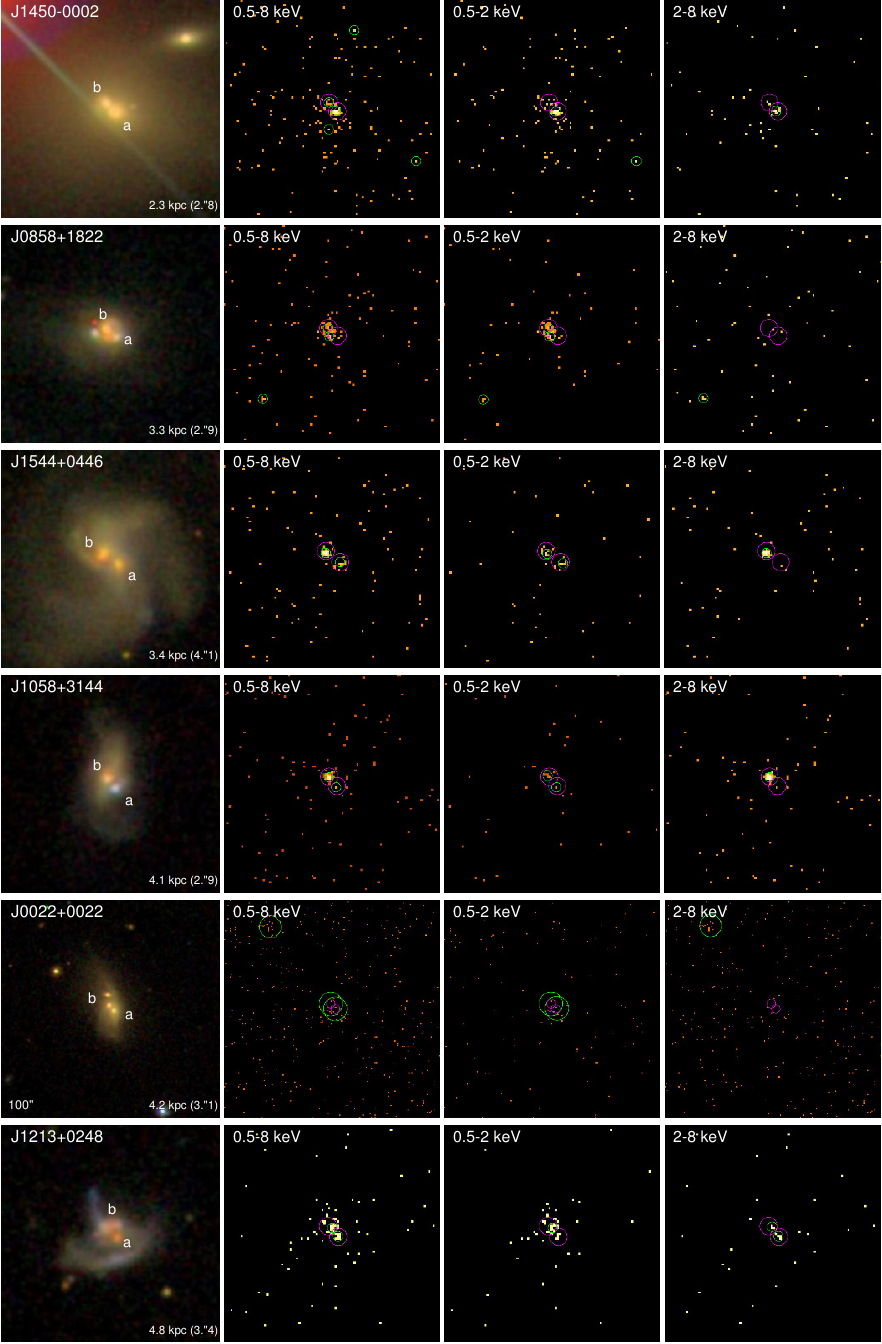}}
\caption{SDSS $gri$-color composite (first column), Chandra 0.5--8 keV (second column), 0.5--2 keV (third column), and 2--8 keV (last column) images of the 21 AGN pairs with both nuclei detected in the X-rays. Each panel is $50\arcsec \times 50\arcsec$ unless otherwise labeled. North is up and east is to the left. The targets are ordered with increasing projected physical separation (as labeled, with angular distance in parenthesis). Magenta circles denote positions of the optical nuclei. Green circles represent the 90\% ECR of the X-ray detections.
}
\label{fig:image_AGNpair}
\end{figure}

\begin{figure}
  \ContinuedFloat 
\centering
\subfloat{\includegraphics[width=0.84\textwidth,angle=0]{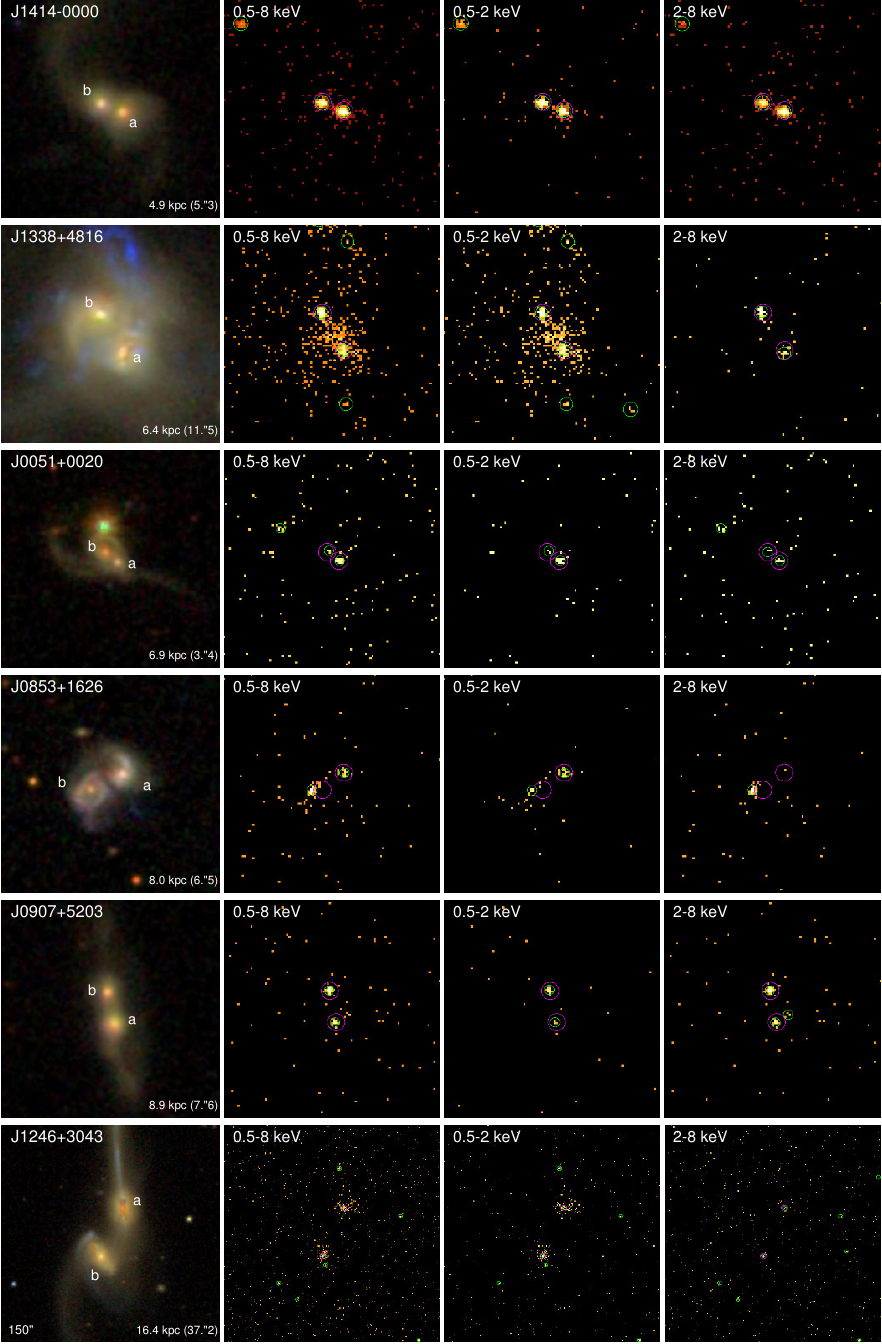}}
\caption{--continued}
\end{figure}

\begin{figure}
  \ContinuedFloat 
\centering
\subfloat{\includegraphics[width=0.84\textwidth,angle=0]{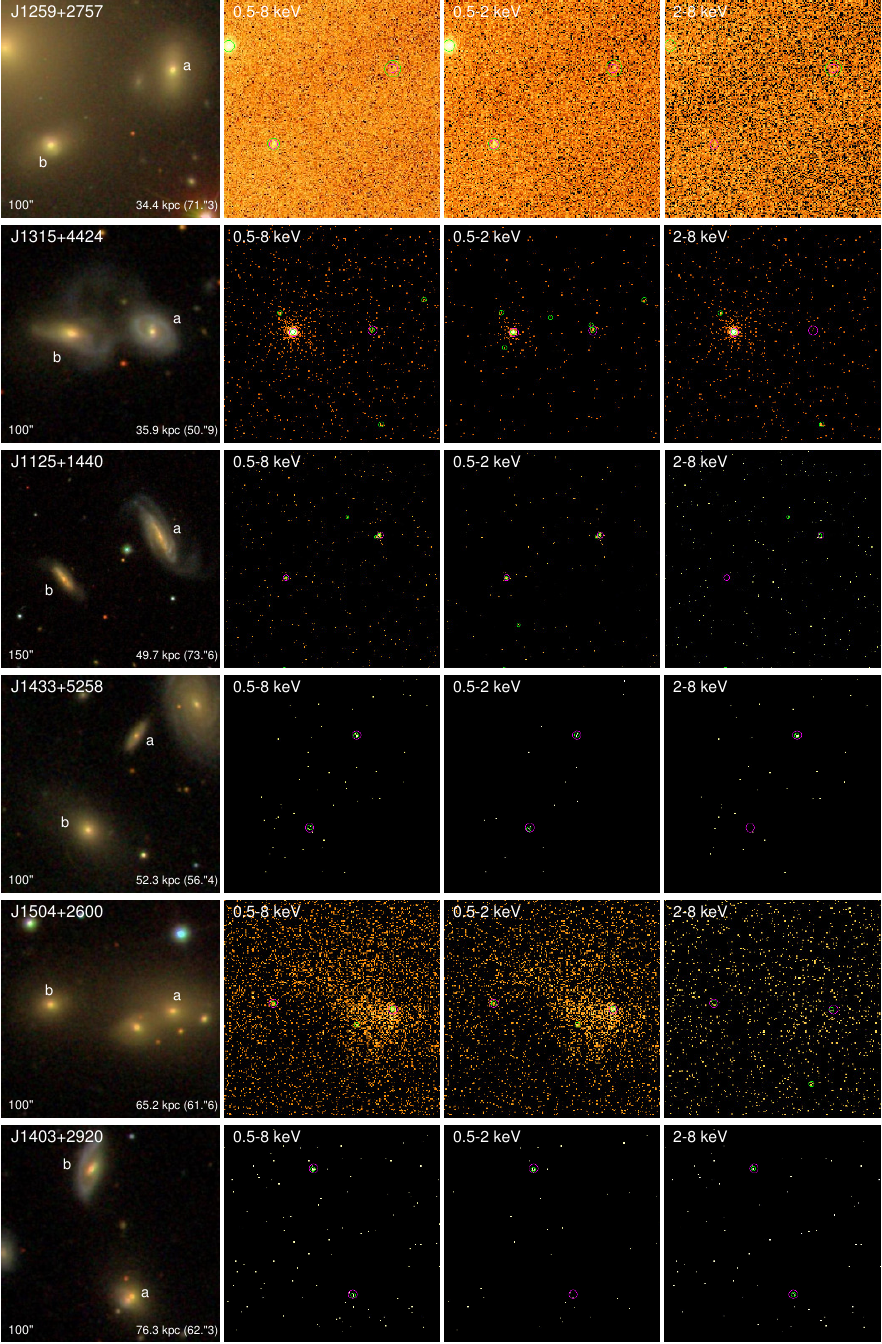}}
\caption{--continued}
\end{figure}

\begin{figure}
  \ContinuedFloat 
\centering
\subfloat{\includegraphics[width=0.84\textwidth,angle=0]{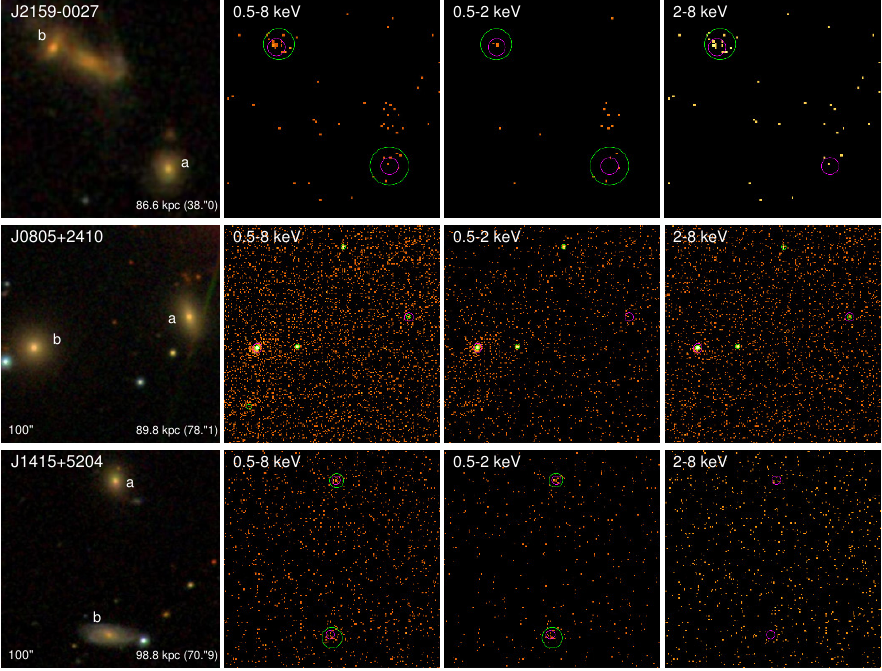}}
\caption{--continued}
\end{figure}

We find an X-ray detection rate of $57\% \pm 7\%$ (66/115) in the control sample of single AGNs in isolated galaxies. There are 60 nuclei detected in the $F$ band, 58 nuclei in the $S$ band, and 36 nuclei detected in the $H$ band. This is consistent with the X-ray detection rate in AGN pairs within uncertainties. Our result suggests that the X-ray properties of optically selected AGN pairs are similar to those of optically selected single AGNs at least at $z{\sim}0.1$.

\begin{figure}\centering
\includegraphics[width=0.9\textwidth,angle=0]{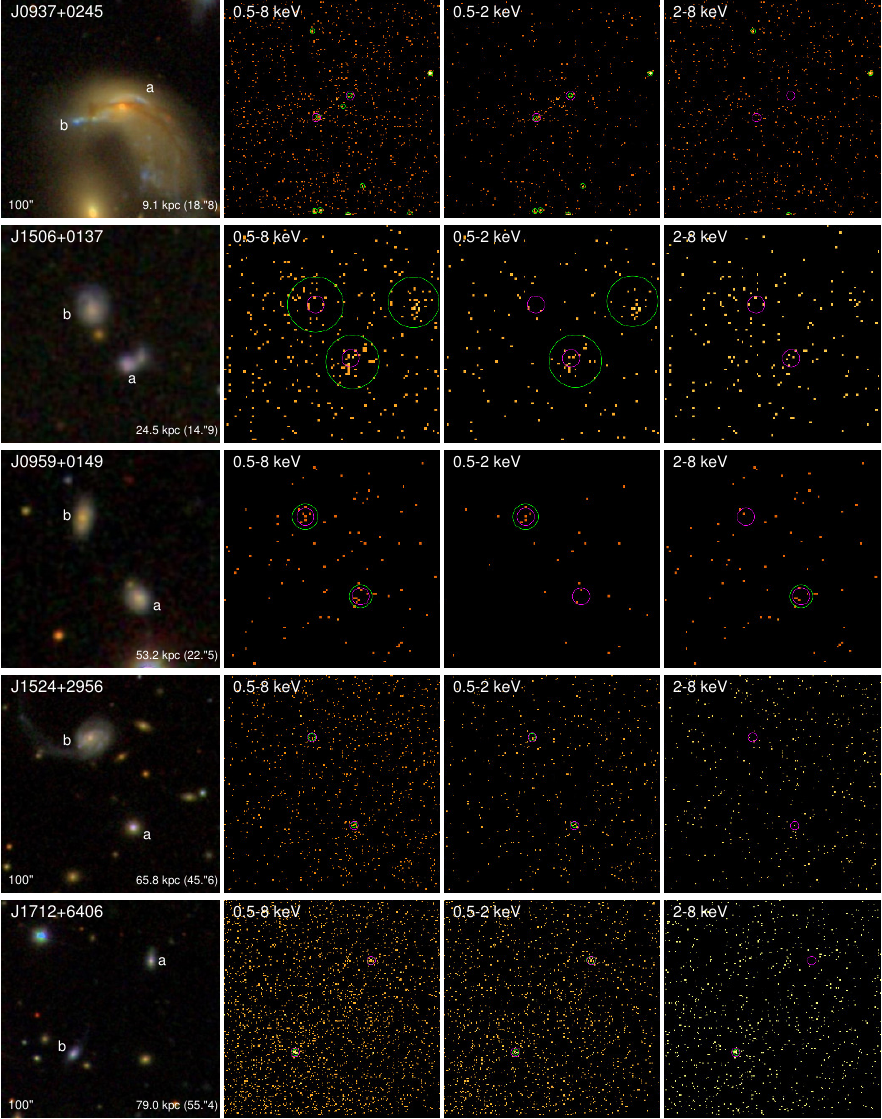}
\caption{
SDSS $gri$-color composite (first column), Chandra 0.5-8 keV (second column), 0.5-2 keV (third column), and 2-8 keV (last column) images of the 5 SFG pairs with both nuclei detected in X-ray. Each panel is $50\arcsec \times 50\arcsec$ unless otherwise labeled. North is up and east is to the left. The targets are ordered with increasing projected separation (as labeled, with angular distance in parenthesis). Magenta circles denote positions of the optical nuclei whereas green circles represent the 90\% ECR of the X-ray detections.
}
\label{fig:image_SFGpair}
\end{figure}

On the other hand, we find a significantly lower X-ray detection rate of $17\% \pm 4\%$ (23/134) in the comparison sample of SFG pairs. There are 22 nuclei detected in the $F$ band, 20 nuclei detected in the $S$ band, and only 6 nuclei detected in the $H$ band. There are 5 SFG pairs with both nuclei detected and 13 pairs with one nucleus detected. Figure~\ref{fig:image_SFGpair} shows the optical and X-ray images of the 5 SFG pairs with both nuclei detected in X-rays. 
For the subsample of 78 nuclei among the SFG pairs with stellar masses $> 10^{9} \rm~M_{\sun}$, which has a more similar stellar mass distribution to that of the AGN pairs, the X-ray detection rate is $21\% \pm 5\%$ (16/78). This is still much lower than that of the AGN pairs ($58\% \pm 7\%$, which is again $58\% \pm 7\%$, or 70/120, if we focus on AGN pairs with stellar masses $> 10^{9} \rm~M_{\sun}$). This indicates that the X-ray emission from the AGN pairs has a minor contribution from star-formation--related processes. Along with the fact that the SFG pairs have a systematically higher SFR distribution than AGN pairs, this significantly lower X-ray detection rate in SFG pairs actually gives an upper limit of the X-ray emission from SF-related processes.
We address X-ray contamination from star-formation--related process further in Section~\ref{subsec:SF}.

The X-ray flux detection limit is highly inhomogeneous and spans three orders of magnitude in our sample, which is primarily caused by the different exposure times of different systems (Table~\ref{tab:info} and Figure~\ref{fig:detlim}). This may introduce selection biases into the resultant X-ray detections. For instance, we would have missed faint sources below the detection limit. We would also have been more likely to detect systems with targeted observations. To quantify this sytematics, we examine the subsamples with an X-ray luminosity detection limit of $\rm~L_{0.5-8,lim} < 10^{41.2}~erg~s^{-1}$. This threshold is chosen based on a two-sample Kolmogorov-Smirnov (K-S) test between the AGN pairs and the control and comparison samples. The significance level (probability of homogeneity) of the K-S statistic is 0.73 between AGN pairs and single AGNs and is 0.46 between AGN pairs and SFG pairs, respectively. This ensures a more homogeneous distribution between different samples and reduces the selection bias. These subsamples have 114 AGN pairs, 94 single AGNs, and 90 SFG pairs. Their X-ray detection rates are $60\% \pm 7\%$ (68/114), $63\% \pm 8\%$ (59/94), and $22\% \pm 5\%$ (20/90). The X-ray detection rate in the AGN pairs remains the same, while those in the single AGNs and SFG pairs increase slightly. Nevertheless, our conclusion on the comparison among different samples still holds. We discuss the implications of the X-ray detection rates in Section~\ref{sec:discussion}.

\subsection{Global X-ray Properties}
\label{subsec:properties}

\begin{figure}\centering
\includegraphics[width=0.43\textwidth,angle=90]{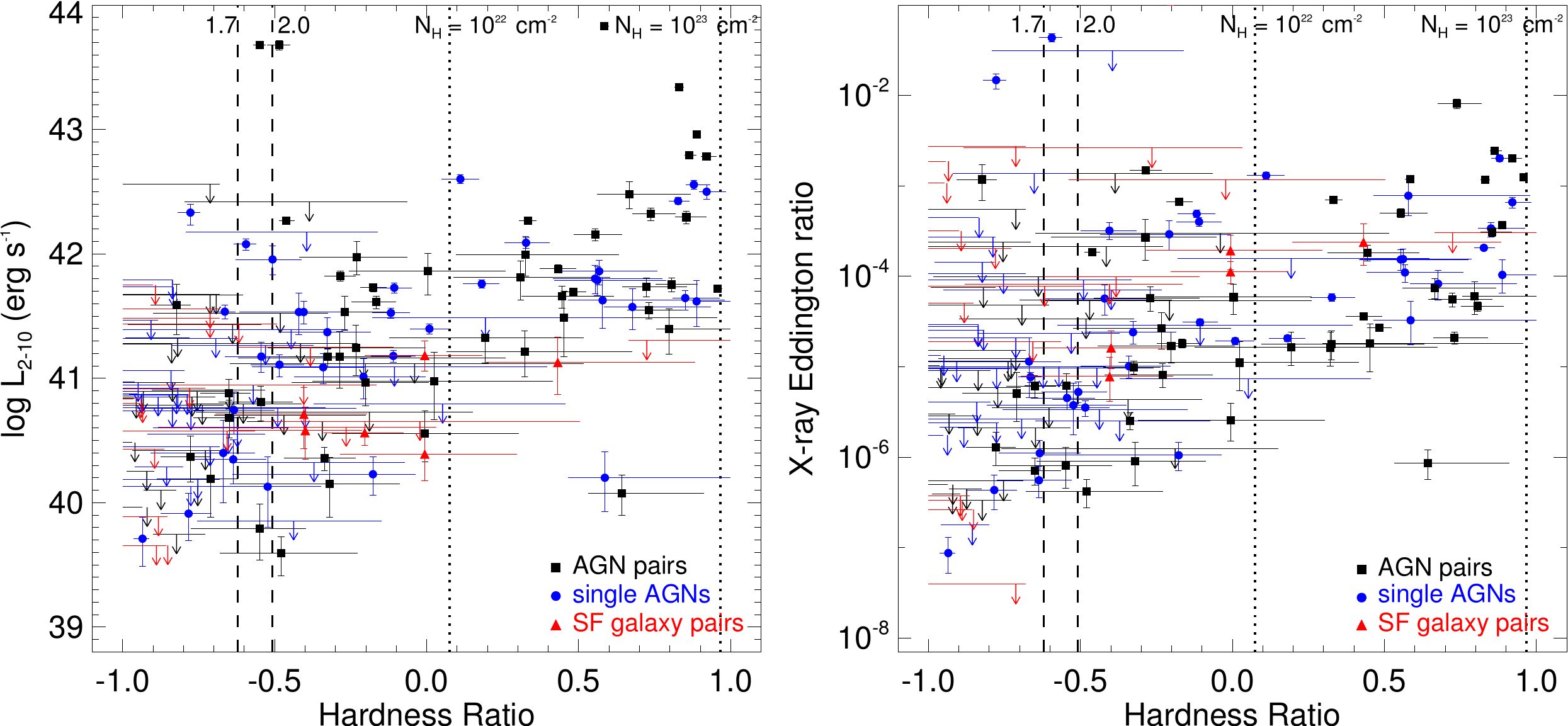}
\caption{{\it Left}: 2--10 keV luminosity vs. hardness ratio. {\it Right}: X-ray Eddington ratio vs. hardness ratio. The black squares, blue circles and red triangles represent X-ray counterparts of AGN pairs, control samples of single AGNs and star-forming galaxy pairs, respectively. The vertical dashed lines, from left to right, correspond to hardness ratios for an absorbed power-law spectrum with a fixed $N_{\rm H} = 10^{21} \rm~cm^{-2}$ and a photon index of 2.0 and 1.7, respectively. The vertical dotted lines correspond to hardness ratios for an absorbed power-law spectrum with $N_{\rm H} = 10^{22}, 10^{23} \rm~cm^{-2}$ and a fixed photon index of 1.7.  
}
\label{fig:hardness}
\end{figure}

The left panel of Figure \ref{fig:hardness} shows the 2--10 keV intrinsic luminosity ($L_{\rm 2-10}$) against hardness ratio of all the X-ray-detected nuclei in the AGN pairs and the control and comparison samples. Table \ref{tab:info} lists the hardness ratio, which is defined as HR = (H - S)/(H + S). It is calculated from the observed counts in the $S$ (0.5--2 keV) and $H$ (2--8 keV) bands using a Bayesian approach \citep{park06}. 
For sources that are not detected in the $H$ band but are detected in the other two bands, we show the 3-$\sigma$ upper limit of $L_{\rm 2-10}$ in the plot. About forty sources, exclusively and about equally from the AGN pairs and single AGNs, are found at the top right portion with high luminosities ($L_{\rm 2-10} \gtrsim 10^{41.2}{\rm~erg~s^{-1}}$) and positive hardness ratios (HR $\gtrsim 0.1$); these are most likely genuine AGNs. Their hardness ratios suggest a moderately high absorption column density of $10^{22-23} {\rm~cm^{-2}}$. 

In contrast, a substantial fraction of sources are found at the bottom left portion with relatively low $L_{\rm 2-10}$ (or $H$ band non-detections) and negative HR; these include the majority of SFG pairs and a similar number of AGN pairs and single AGNs, indicating that the X-ray emission of these nuclei are more likely dominated by star formation activity rather than an AGN. However, it is still possible that some of these soft, under-luminous sources host an AGN in the center, but the hard X-ray emission is totally obscured due to a very high column density (i.e., Compton thick), leaving only circumnuclear soft X-ray emission falling within the {\it Chandra} aperture. 

There are very few classical AGNs found among all the X-ray-detected nuclei, with all but four sources having $L_{\rm 2-10} > 10^{43}{\rm~erg~s^{-1}}$. 
This is further evidenced by the right panel of Figure \ref{fig:hardness}, which plots the X-ray Eddington ratio against hardness ratio for sources that have a reliable estimate of the black hole mass (inferred from $\sigma_{\ast}$ assuming the $M_{{\rm BH}}$-$\sigma_{\ast}$ relation from \citealp{Gultekin2009}). 
We have checked the SDSS spectra and verified that the $\sigma_{\ast}$ is generally well measured in our sample, at least for pairs still at large projected separations.
Only four X-ray nuclei, one from AGN pairs 
and three from single AGNs, have an X-ray Eddington ratio of $\gtrsim 1\%$.

\subsection{X-ray Contribution from Star Formation}
\label{subsec:SF}

Even in the absence of an AGN, circumnuclear star formation can produce copious X-ray emission, primarily attributed to high-mass X-ray binaries and diffuse hot gas heated by supernovae. We use independent SDSS spectroscopic star formation rates (SFR) given by the MPA-JHU DR7 catalog (listed in Table \ref{tab:info}) to estimate the X-ray luminosity contributed from star formation. We adopt the empirical relation of \citet{ranalli03} given by, 
\begin{equation}\label{eq:lxs_sfrs}
L^{{\rm SF}}_{0.5-2} = 4.5 \times 10^{39} \frac{{\rm SFR}}{M_{\odot}~{\rm
yr}^{-1}} {\rm erg~s}^{-1}, \\
\end{equation}
\begin{equation}\label{eq:lxh_sfrh}
L^{{\rm SF}}_{2-10} = 5.0 \times 10^{39} \frac{{\rm SFR}}{M_{\odot}~{\rm
yr}^{-1}} {\rm erg~s}^{-1},
\end{equation}
which has an rms scatter of 0.27 dex and 0.29 dex in the 0.5--2 keV and 2--10 keV band, respectively.

\begin{figure}\centering
\includegraphics[width=0.43\textwidth,angle=90]{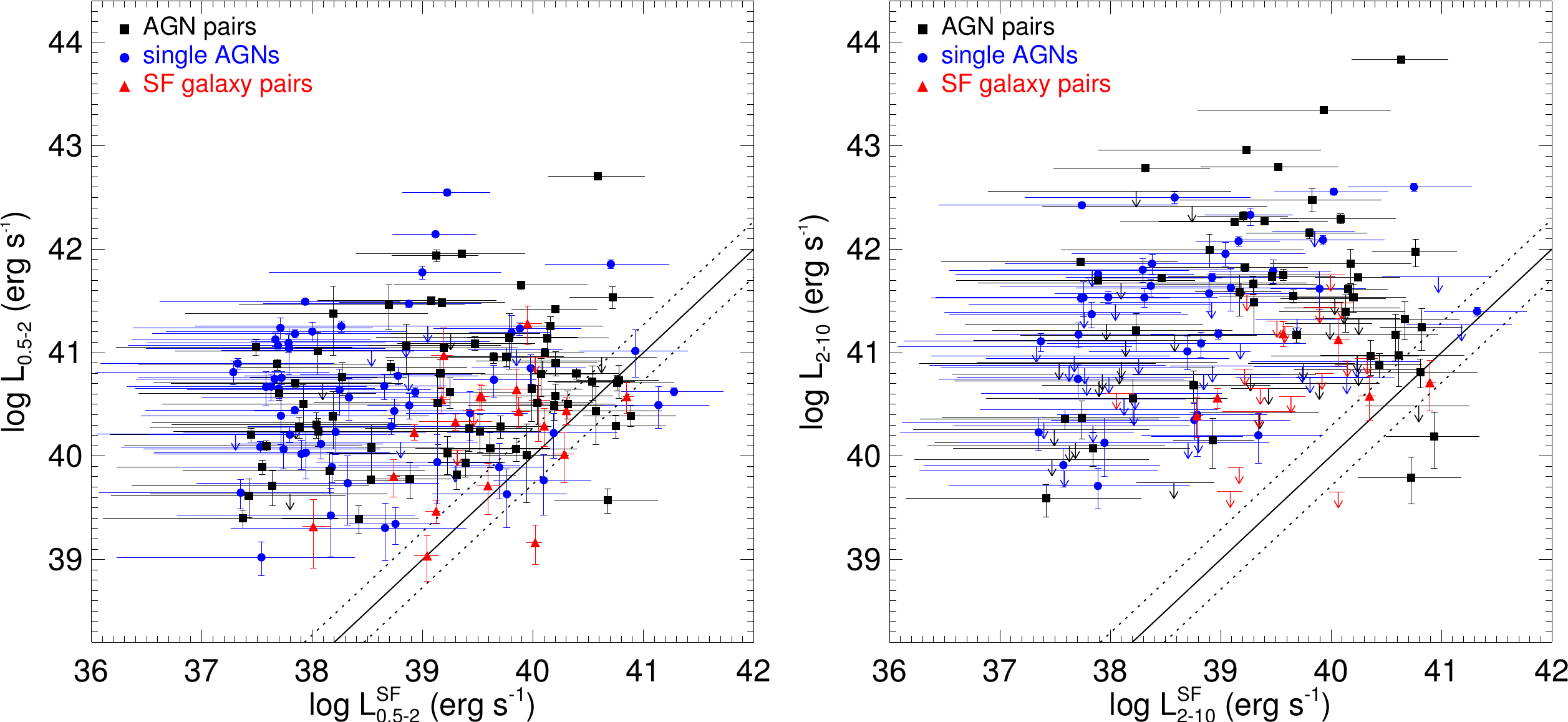}
\caption{Measured 0.5-2 keV ({\it left}) and 2--10 keV ({\it right}) luminosity vs. predicted luminosity due to star formation. The black squares, blue circles and red triangles represent X-ray counterparts of AGN pairs, control samples of single AGNs and star-forming galaxy pairs, respectively. The black solid line indicates a 1:1 relation, with the pair of dashed lines representing the rms scatter.
}
\label{fig:lxsfr}
\end{figure}

Figure \ref{fig:lxsfr} shows the comparison between the measured X-ray luminosity and the empirical X-ray luminosity due to star formation in the two bands. The left panel shows that the majority of AGN pairs and single AGNs are located above the empirical relation, indicating that star formation contributes little to the observed 0.5--2 keV emission from these nuclei. 

However, there is still a significant fraction of AGN pairs overlapping with the SFG pairs, with a notable fraction of the latter sitting above the empirical relation. This may suggest either an underestimate of star formation contribution by the calibration of \citet{ranalli03} (e.g., due to a redshift-dependent effect),
or the presence of an optically hidden AGN that moderately contributes to the soft X-ray emission.

The right panel shows that the predicted contribution from star formation for almost all AGN pairs (as well as single AGNs) is well below the measured 2--10 keV luminosity. The SFG pairs are well separated from the AGN pairs, with only few SFG pairs firmly detected in this band. We have also compared the prediction of \citet{ranalli03} with the empirical correlation between 2--10 keV luminosity $\rm L^{gal}_{HX}$, SFR, and stellar mass from \citet{Lehmer2010}, given by 
\begin{equation}
L^{{\rm gal}}_{{\rm HX}} = \alpha M_{*} + \beta {\rm SFR},
\end{equation}
where $\alpha=(9.05\pm0.37)\times10^{28}$ erg s$^{-1}$ $M_{\odot}^{-1}$ and $\beta=(1.62\pm0.22)\times10^{39}$ erg s$^{-1}$ ($M_{\odot}$ yr$^{-1}$)$^{-1}$. The calibration was based on a sample of 17 LIRGs. The X-ray emission includes both the contribution from high-mass X-ray binaries and low-mass X-ray binaries. The predicted 2--10 keV luminosities are consistent within uncertainties between AGN pairs and SFG pairs.
This suggests that the 2--10 keV emission of AGN pairs (as well as single AGNs) is dominated by genuine AGN activity. Single AGNs tend to have a lower SFR than the AGN pairs, which may indicate that the X-ray emission from single AGNs are even less effected by star formation. 

In view of the above, we consider X-ray nuclei with $L_{2-10} > 10^{41}{\rm~erg~s^{-1}}$ as genuine AGNs. In this more conservative case, the X-ray detection rate is $27\% \pm 4\%$ for AGN pairs, $24\% \pm 5\%$ for single AGNs, and $1\% \pm 1\%$ for SFG pairs (Table 3). Below we focus our discussion on the 2--10 keV band. 



\subsection{Comparison with \OIII\ Luminosity}
\label{subsec:xrayoiii}

\begin{figure}\centering
\includegraphics[width=0.43\textwidth,angle=90]{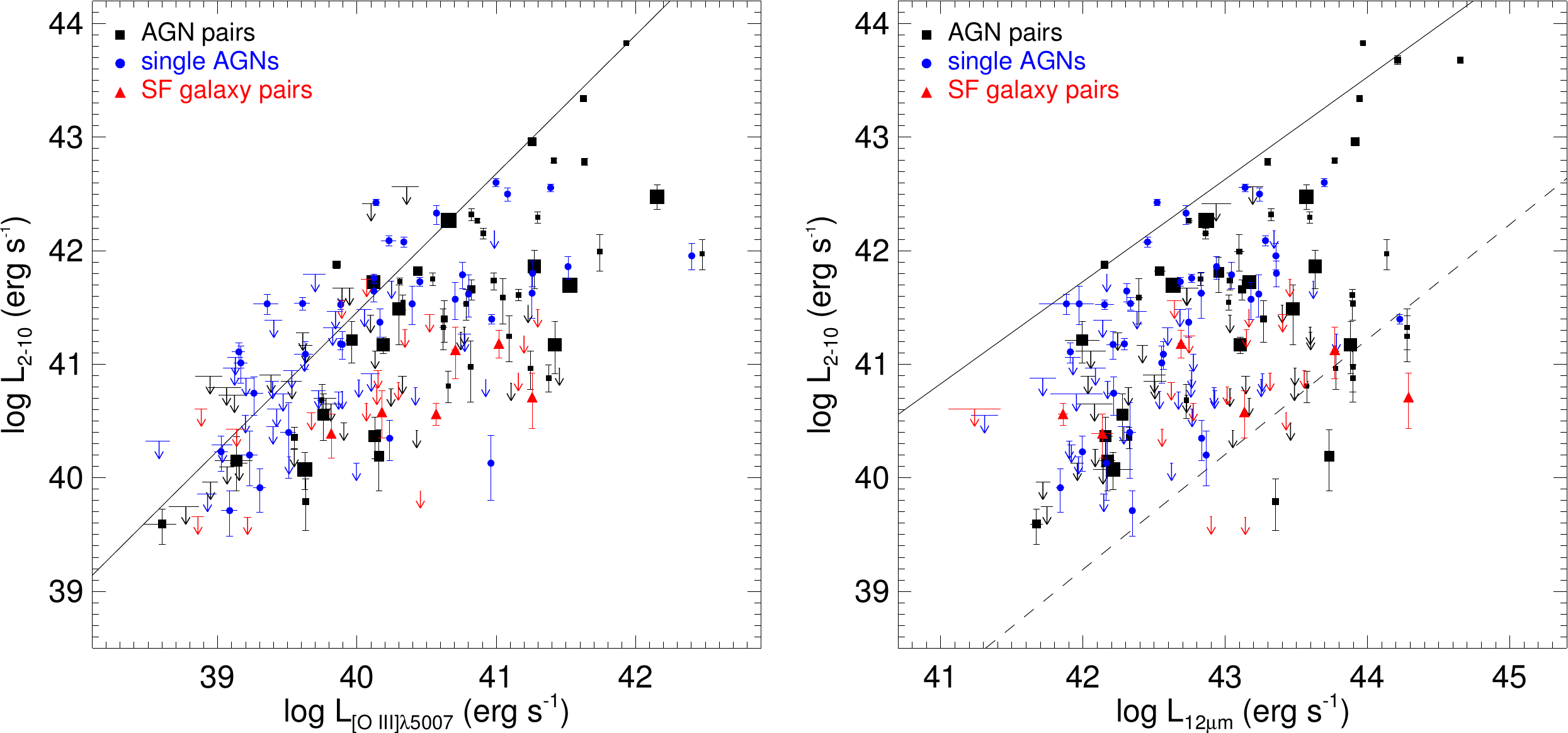}
\caption{{\it Left}: Measured 2--10 keV luminosity vs. extinction-corrected \OIII\ luminosity. The black solid line indicates the mean relation for optically selected Type 1 and Type 2 Seyferts \citep{panessa06}
{\it Right}: Measured 2--10 keV luminosity vs. mid-infrared 12 $\micron$ luminosity.  The black solid line indicates the mean relation for local Seyferts \citep{Gandhi2009}, while the dashed line is a predicted relation between the two quantifies if both are related to star formation (see text).
In both panels, the black squares, blue circles and red triangles represent X-ray counterparts of AGN pairs, control samples of single AGNs and star-forming galaxy pairs, respectively. The size of the black squares is linearly proportional to the projected separation ($r_p$) of the pair in which a given AGN resides.
}
\label{fig:lxo3}
\end{figure}

The left panel of Figure \ref{fig:lxo3} shows the 2--10 keV luminosity against the extinction-corrected \OIII\ luminosity. The extinction of \OIII\ luminosities has been corrected according to the observed Balmer decrement.
The \OIII\ luminosity is an empirical AGN indicator, which is related to the 2--10 keV X-ray luminosity as \citep{panessa06},
\begin{equation}
\log L_{{\rm 2-10\,keV}} = (1.22\pm 0.06)\log L_{{\rm [O\,{\tiny III}]}} + (-7.34 \pm 2.53). 
\end{equation}
The relation was calibrated based on both optically-selected Type 1 and Type 2 Seyferts after extinction correction. Objects in our sample of single AGNs are consistent with the empirical relation at least over the range of $10^{39}{\rm~erg~s^{-1}} \lesssim L_{{\rm [O\,{\tiny III}]}} \lesssim 10^{41}{\rm~erg~s^{-1}}$, where the majority of single AGNs are found. On the other hand, the AGN pairs on average lie below the empirical relation, which indicates a systematically smaller X-ray-to-\OIII\ luminosity ratio in AGN pairs than in single AGNs in isolated galaxies. 

Previous work based on smaller samples has found evidence for a systematically smaller hard-X-ray-to-\OIII\ luminosity ratio in optically selected dual AGNs than in single AGNs \citep[e.g.,][]{Liu2013,Hou2019}. 
With a much larger sample of AGN pairs, we here confirm the X-ray weak tendency in optically selected AGN pairs. This puts the result on firmer statistical ground. 
The low-X-ray-to-\OIII\ ratio may be due to the \OIII\ emission predominately arising from the narrow-line region that is subject to less obscuration that the X-rays arising from the immediate vicinity of the SMBH. 
Moreover, among the $\sim$60 nuclei in AGN pairs found below the empirical line, roughly half have $r_p \lesssim$ 10 kpc and half have $r_p \gtrsim$ 10 kpc. 
Finally, the SFG pairs are almost exclusively located below the AGN relation, which is consistent with the picture where star-formation--related processes produce systematically weaker X-ray emission at a given \OIII\ luminosity.

\subsection{Comparison with Mid-Infrared Luminosity}
\label{subsec:xraymir}

The right panel of Figure \ref{fig:lxo3} compares the 2--10 keV luminosity with the mid-infrared (MIR) 12 $\mu$m luminosity. The 12 $\mu$m luminosity was derived from the {\it WISE} survey \citep{Wright2010}. Shown for context is an empirical relation (the solid line) between the X-ray and MIR luminosities of a sample of local Syferts of both Type 1 and Type 2 \citep{Gandhi2009} given by, 
\begin{equation}
\log (L_{\rm 12{\mu}m}/10^{43}{\rm~erg~s^{-1}}) = (0.19 \pm 0.05) + (1.11 \pm 0.07) \log (L_{2-10}/10^{43}{\rm~erg~s^{-1}}).
\end{equation}
The MIR emission is thought to arise from the torus reprocessing the AGN radiation. 

Nearly all AGN pairs fall below the empirical relation found in single AGNs, by up to $\sim$2 dex for a given 12 $\mu$m luminosity. This appears to suggest that our 2--10 keV luminosity is significantly underestimated, e.g., due to strong obscuration caused by a merger-driven gas concentration effect. However, we argue that this is unlikely for two reasons. First, the observed X-ray hardness ratio suggests that only a small number of nuclei in AGN pairs have an absorption column density close to $10^{23} {\rm~cm^{-2}}$ (Figure~\ref{fig:hardness}), at which level the 2--10 keV luminosity would only be decreased by $\sim$0.4 dex \citep[e.g.,][]{Satyapal2017}. Second, there is no systematic difference between the X-ray hardness ratio distributions of AGN pairs and single AGNs, which indicates no systematic excess in the X-ray obscuration in AGN pairs relative to single AGNs. 

A more plausible explanation is an excess MIR luminosity due to contamination by the host galaxy, since the {\it WISE} aperture (FWHM $6\arcsec-12\arcsec$) is significantly larger than the {\it Chandra} PSF. This is supported by the fact that the SFG pairs have similar 12 $\mu$m luminosities to those seen in the AGN pairs. For instance, a typical SFR of $10{\rm~M_\odot~yr^{-1}}$ can account for a 12 $\mu$m luminosity of $10^{43.5}{\rm~erg~s^{-1}}$, according to the $L_{\rm 12{\mu}m}$-SFR relation of \citet{Donoso2012}, 
\begin{equation}
\log (L_{\rm 12{\mu}m}^{\rm SF}/{\rm erg~s^{-1}}) = 0.987\log ({\rm SFR}/{\rm M_\odot~yr^{-1}}) + 42.5. 
\end{equation}
This can be further translated into a $L_{\rm 2-10}^{\rm SF} - L_{\rm 12{\mu}m}^{\rm SF}$ relation assuming that both luminosities are due to star formation \citep{Gross2019}, which is shown by the dashed line in the right panel of Figure \ref{fig:lxo3}. 
Those AGN pairs with $L_{\rm 2-10} \lesssim 10^{41}{\rm~erg~s^{-1}}$ may suffer from an overestimation of the AGN MIR luminosity due to contamination from star formation.
We have also checked the {\it WISE} $W1$ ($3.4{\mu}m$) - $W2$ ($4.6{\mu}m$) colors of our sample. Most of AGN pairs have $W1 - W2 \sim 0.1$, which are typical of spiral galaxies (\citealt{Wright2010}, Figure 12 therein). Only 25 nuclei have $W1 - W2 > 0.5$, which can be regarded as classical AGN. This lends further support to our suggestion that the MIR luminosity of most AGNs is contaminated by host galaxies because of the much larger PSF of {\it WISE} than {\it Chandra}. 

\subsection{AGN X-ray Luminosity versus Projected Separation}
\label{subsec:lxrp}


\begin{figure}\centering
\includegraphics[width=0.44\textwidth,angle=90]{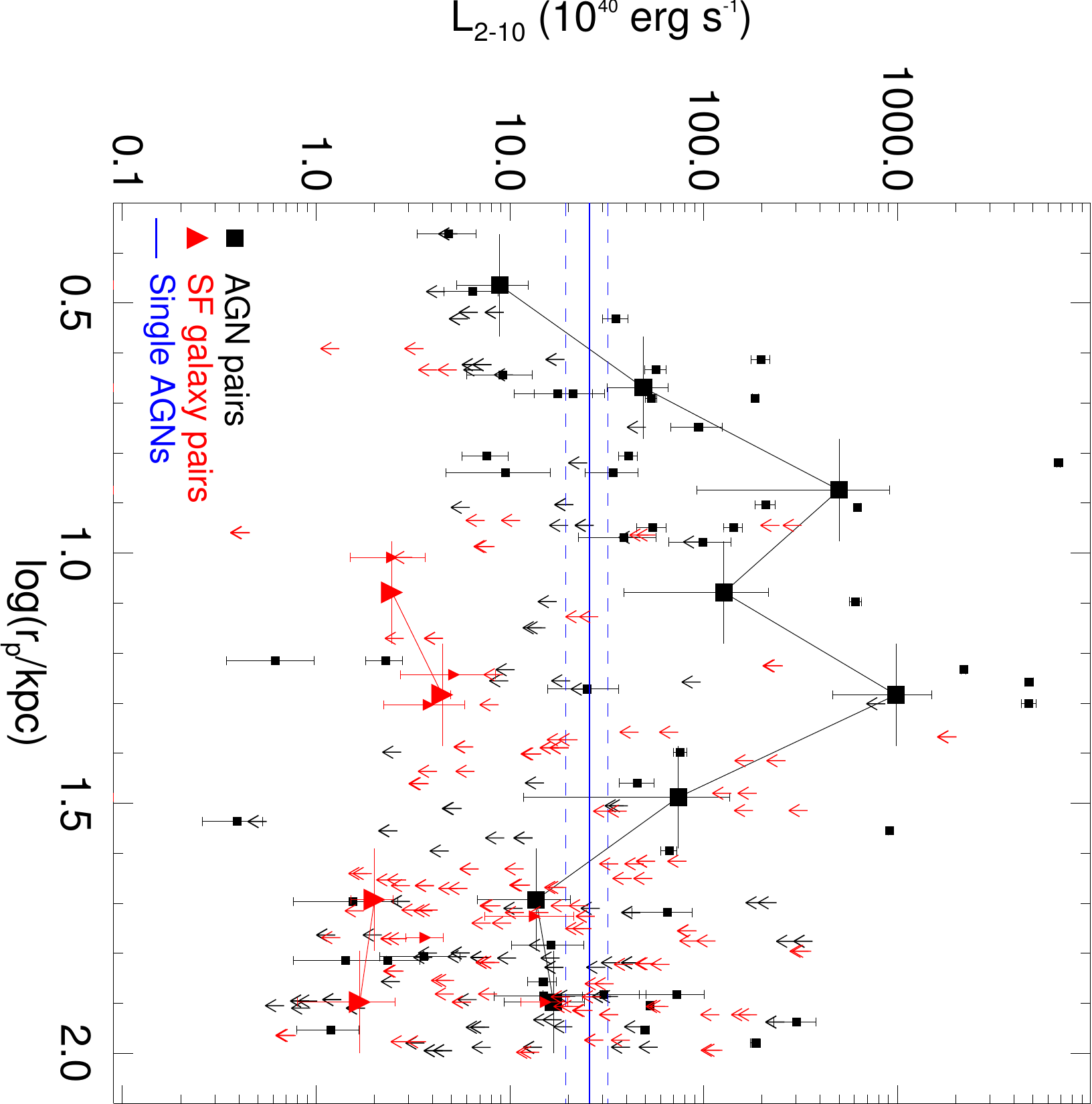}
\includegraphics[width=0.44\textwidth,angle=90]{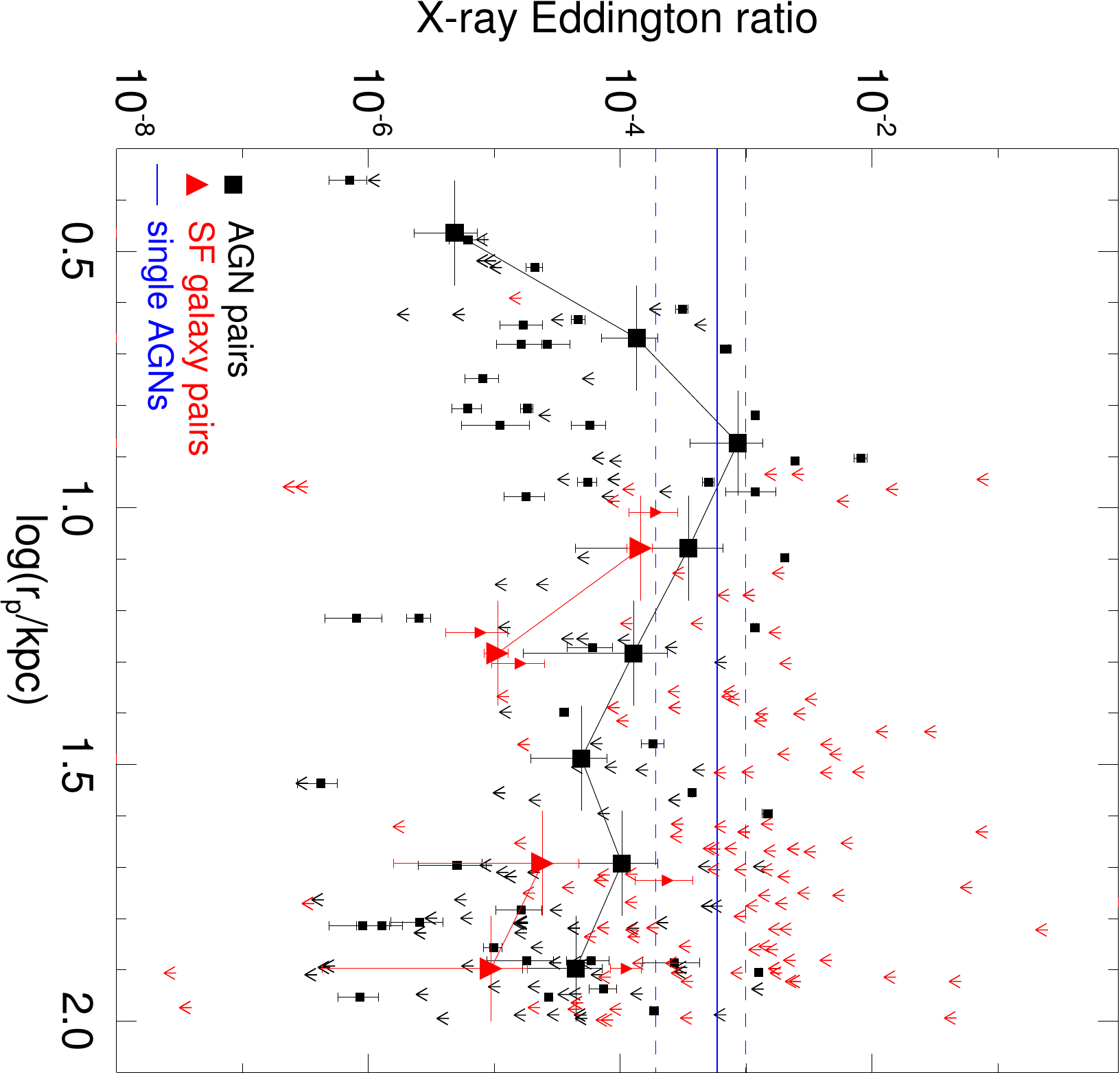}
\caption{X-ray 2--10 keV luminosity ({\it left}) and X-ray Eddingtion ratio ({\it right}) as a function of projected physical separations $r_p$. The X-ray detected AGN pairs and star-forming galaxy pairs are shown in smaller black squares and red triangles, respectively. The 3-$\sigma$ upper limit of undetected nuclei are shown in arrows. For each $r_p$ bin, the mean luminosities of AGN pairs and star-forming galaxy pairs are shown in larger black squares and red triangles with $1\sigma$ error bars connected by line, respectively. 
In a few bins the mean luminosity of star-forming galaxy pairs is undetermined due to the lack of firm detections in these bins. The blue horizontal solid line shows the mean value of single AGNs, and the dashed blue lines represent the $\pm 1\sigma$ error bars.
}
\label{fig:lxrp}
\end{figure}

The sizable sample of AGN pairs allows us, for the first time from an X-ray perspective, to examine the causal link between AGN activity and pair separation, which has been suggested by simulations of galaxy merger \citep[e.g.,][]{Rosas-Guevara2019}. The projected pair separation is used as a proxy for the merger phase, where larger (smaller) $r_{p}$ represents the earlier (later) stage of a merger. 

Figure \ref{fig:lxrp}a shows the 2--10 keV luminosity or upper limits for non-detected sources as a function of projected separations $r_p$ for AGN pairs and the comparison sample of SFG pairs. Perhaps the most surprising finding is the paucity of X-ray luminous nuclei at small separations ($r_p \lesssim$ 4 kpc). 
Moreover, a clear trend is revealed when we bin the AGN pairs by their separations to reduce the scatter in the mean. The mean X-ray 2--10 keV luminosity of each $r_p$-bin is calculated using Astronomy SURVival Analysis \citep[ASURV;][]{Feigelson1985}, which takes both detections and upper limits into account. For targets with $r_p \lesssim 20$ kpc, the mean X-ray 2--10 keV luminosity declines as $r_p$ decreases. This trend is the opposite to that seen in the \OIII\ luminosity versus $r_p$ relation in AGN pairs\citep[][see also below]{Liu2012} and that seen in single AGNs in galaxy pairs \citep[e.g.,][]{ellison11}. The mean X-ray 2--10 keV luminosity of the innermost bin ($r_p \lesssim 4$ kpc) is significantly lower than the mean luminosity averaged over all single AGNs (also calculated with ASURV; represented by the blue lines in Figure \ref{fig:lxrp}). The mean X-ray luminosity also decreases as $r_p$ increases from $\sim$20 kpc beyond until about 100 kpc, which is the largest separation spanned by our samples.
We should note that the particularly high third and fifth bins are partially due to three most luminous nuclei with $L_{\rm X,2-10} > 4 \times 10^{43}{\rm~ erg~s^{-1}}$. But even without these three nuclei, the overall trend remains qualitatively unchanged.
Figure \ref{fig:lxrp}b shows the X-ray Eddington ratio as a function of projected separations $r_p$ for AGN pairs and the comparison sample of SFG pairs.
Only nuclei with measured black hole mass are included in the figure. The overall trend is consistent with that in Figure \ref{fig:lxrp}a but the peak moves to $\sim$8 kpc.

We have tested systematic effects the inhomogeneous detection limits by including only those pairs with a detection limit of $\rm~L_{0.5-8,lim} < 10^{41.2}~erg~s^{-1}$ (see also Section~\ref{subsec:detectionrate}). The overall trend stays unaffected using this homogeneous detection limit. 

For the SFG pairs, due to the lack of X-ray detected nuclei (6 sources) particularly at $r_p \lesssim 4$ kpc, we only have four bins with available mean 2--10 keV luminosities. Nevertheless, the trend in the SFG pair sample is still generally flat. The mean X-ray 2--10 keV luminosities are much lower than those in both AGN pairs and single AGNs. 

\begin{figure}\centering
\includegraphics[width=0.5\textwidth,angle=90]{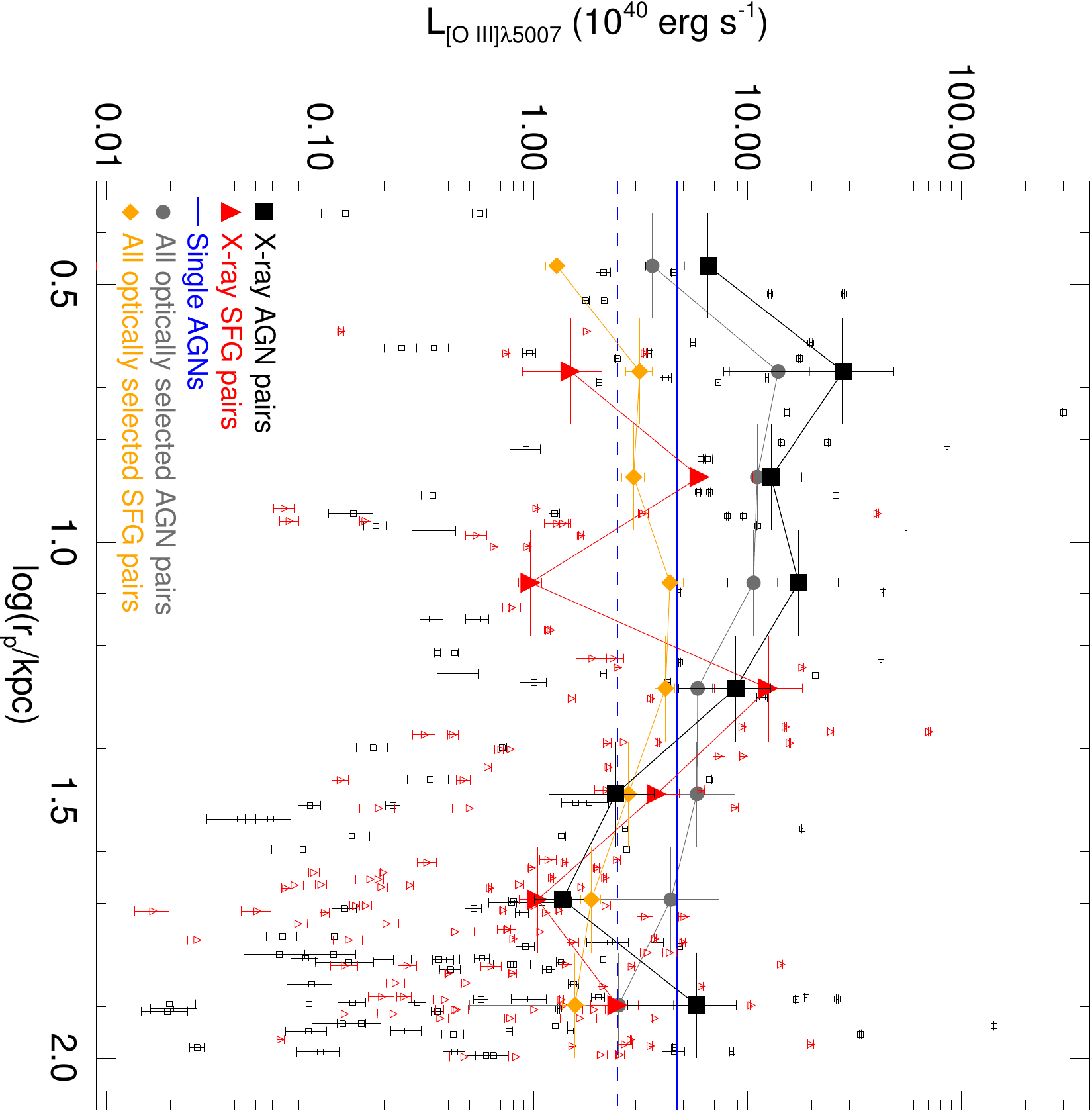}
\caption{\OIII\ luminosity as a function of projected separations $r_p$. All AGN pairs and star-forming galaxy pairs with {\it Chandra} observation are shown in smaller open black squares and red triangles, respectively. For each $r_p$ bin, the mean luminosities of AGN pairs and star-forming galaxy pairs are shown in larger solid black squares and red triangles connected by line, respectively. We also show the mean luminosities of all optically selected AGN pairs in gray solid circles and the mean luminosities of all optically selected star-forming galaxy pairs in orange solid diamonds. 
The blue horizontal solid line shows the mean value of \OIII\ luminosity of single AGNs.
}
\label{fig:lo3rp}
\end{figure}

Figure \ref{fig:lo3rp} shows the mean \OIII\ luminosity as a function of projected separation $r_p$. We use the mean value to be consistent with the above analysis of the X-ray luminosity. For all AGN pairs with {\it Chandra} observation, the mean \OIII~luminosity increases with decreasing separation. This increasing trend extends until the innermost bin at $r_p \sim$ 4 kpc, which is much more inward than that seen in the X-ray luminosity. The separation dependence flattens out at large $r_p$ ($\gtrsim$30 kpc). There was no declining trend in the innermost bin in a similar plot of \citet{Liu2012}. There are two possible reasons for this difference. First, the \OIII\ luminosity quoted in \citet{Liu2012} has been corrected for contribution from star formation following the procedure of \citet{kauffmann09}. Second, we quote mean values here whereas median values were quoted in \citet{Liu2012} and the difference may be due to small number statistics. The innermost bin is highly incomplete due to the angular resolution limit of SDSS imaging ($\sim$2$''$ corresponding to $\sim$4 kpc at redshift $z{\sim}0.1$).


\section{Discussion}\label{sec:discussion}

The current work represents the first to study the effects of mergers in optically selected AGN pairs from an X-ray perspective in a statistically meaningful sense. Previous studies based on the parent sample of optically-selected AGN pairs has found evidence for tidally enhanced AGN in both AGN pairs \citep[e.g.,][]{Liu2012} and in single AGNs in galaxy pairs \citep[e.g.,][]{ellison11}. These studies, however, have adopted the \OIII~luminosity as an indicator for AGN luminosity. As discussed in Section~\ref{sec:intro}, the X-ray band provides not only an independent but also a more direct and robust diagnostic of AGN. This is especially helpful for the regime of low luminosity/low Eddington ratio characteristic of the current sample, where the casual link between mergers and AGN has been controversial and more difficult to probe due to the competition of secular processes \citep[e.g.,][]{Liu2012}.  

Our X-ray census enriches this picture in two aspects. First, the X-ray detection rate in AGN pairs is substantially higher than that in SFG pairs, which holds true for the whole sample as well as several sub-samples constructed to test various systematics (Section~\ref{subsec:detectionrate} and Table \ref{tab:rate}). In the more conservative case where a genuine AGN is required to have an X-ray 2--10 keV luminosity greater than $10^{41}{\rm~erg~s^{-1}}$, the detection rate is $\sim$27\%. The X-ray detection rate provides a useful constraint for the triggering of AGN pairs, which is the subject of several recent theoretical studies \citep{Capelo2017, Rosas-Guevara2019,Solanes2019,YangC2019}. For instance, using a suite of idealized major merge simulations, \citet{Solanes2019} predicted an abundance of $5\%-30\%$ weakly accreting AGN pairs (i.e., with bolometric luminosities $\lesssim10^{43}{\rm~erg~s^{-1}}$) under an observational condition tailored to match the selection criteria of our parent sample \citep{Liu2011a}. While current simulations generally cannot self-consistently predict the accretion-induced X-ray luminosity, our X-ray detection rates appear to be broadly consistent with predictions from simulations.

Second, the X-ray detection rate is statistically indistinguishable between the samples of AGN pairs and single AGNs in isolated galaxies. Moreover, the samples of single AGNs and AGN pairs span a similar range in X-ray luminosity and Eddington ratio (Figure~\ref{fig:hardness}). This underscores the challenge in distinguishing the roles of galaxy-galaxy interaction and internal, secular processes in triggering and maintaining a moderate level of SMBH accretion. Our result is consistent with the picture where the overall weak enhancement in AGN luminosity/Eddington ratio observed in optically-selected close AGN pairs represents enhanced BH accretion on top of a background of low-level AGN driven by secular processes not directly associated with galaxy interactions \citep{Liu2012}.

Nevertheless, the mean 2--10 keV luminosity as a function of projected separation shown in Figure \ref{fig:lxrp} does suggest that galaxy-galaxy interactions do affect the X-ray emission.
The increase in the X-ray luminosity as $r_{p}$ decreases from $r_{p} \gtrsim$ 50 kpc to $r_{p} \sim 20$ kpc may be evidence for enhanced SMBH accretion at the first pericentric passage at the early stage of a merger, consistent with predictions from simulations \citep[e.g.,][]{Rosas-Guevara2019}. The decrease of the X-ray luminosity from $r_{p} \sim$ 15 kpc inward, however, is rather surprising. This appears to be contradictory to expectations from simulations, where tidal torques would funnel more gas into the center to drive stronger SMBH accretion as observed in \OIII\ luminosity \citep[e.g.,][]{Liu2012}. 

There are two possible explanations for the decrease of X-ray luminosity at the smallest separations. First, merger-induced gas inflows could become so strong that a central concentration of cold gas heavily obscures even the hard (2--10 keV) X-rays. Indeed, there is growing evidence for heavily obscured close AGN pairs on kpc-scales \citep[e.g.,][]{Satyapal2017,Pfeifle2019}. The systematically lower X-ray-to-\OIII\ luminosity ratio in the AGN pairs (Section~\ref{subsec:xrayoiii}) may also be a merger-driven gas concentration effect. This would preferentially obscure the X-rays which originate from regions closer to the SMBH than the \OIII\ emission. However, a low X-ray-to-\OIII\ luminosity ratio is found in AGN pairs of both small and large separations. Therefore, other factors such as a selection bias in the \OIII\ luminosity cannot be completely ruled out \citep[e.g.,][]{Liu2013}. 
We also note that the empirical \OIII\ and X-ray luminosity relation has a large scatter. Especially for galaxies with strong star formation activities, the \OIII\ luminosity would not be a accurate indicator for AGN luminosity. 

Second, AGN feedback triggered by the gas inflow could act to heat and subsequent expel gas from the central region and suppress further substantial SMBH accretion \citep{Fabian2012}. Our results may be compared with future merger simulations that incorporate AGN feedback, which can shock heat the interstellar medium and drive strong outflows. The effect of feedback in AGN pairs can in principle be tested using direct tracers such as the circumnuclear diffuse hot gas and/or outflows of ionized gas, which had been observed in merger systems \citep[e.g.,][]{Greene2014}. In fact, in an appreciable fraction of AGN pairs studied here, we find clear evidence for the presence of circumnuclear diffuse X-ray emission (such as in J1213+0248, J1338+4816 (i.e., Mrk266), J0853+1626, J1504+2600, etc. in Figure \ref{fig:image_AGNpair}), which may be related to AGN feedback. We defer a systematic study of diffuse X-ray emission around AGN pairs to future work. 


Finally, we discuss the caveats and limitations of our results. While representing the currently largest sample of X-ray-detected AGN pairs matched from an optically selected sample, 
it is still subject to small number statistics. 
Our AGN pairs sample consists of 67 pairs (134 nuclei) and each bin in the X-ray luminosity trend as a function of projected separation consists of $\sim 10-15$ nuclei.
Future X-ray observations with much larger samples are needed to put our results on firmer statistical ground. The current sample is also strongly biased toward AGN pairs with close separations (20 of the 67 pairs have $r_{\rm p} \lesssim$ 10 kpc), due to the fact that essentially all targeted {\it Chandra} observations were for close pairs. Only $\sim$70\% of our archival sample were observed by chance and would be free from a selection bias. Fortunately, this selection bias toward small $r_p$ actually helps reveal the paucity of X-ray-luminous nuclei in close pairs. A complete study of the effects of merger-driven AGN must also include interacting galaxy pairs at large separations (10 kpc $\lesssim r_{\rm p} \lesssim$ 100 kpc), which should be more favorably targeted by future observations.

In zoom-in merger simulations, a galaxy pair would evolve in isolation, i.e., free from perturbation by a third galaxy, which is not necessarily the case in the observations. The comparison may be improved by carefully choosing truly isolated galaxy pairs for future observations \citep[e.g.,][]{Argudo-Fernandez2015}. Another mismatch between the current {\it Chandra} sample and merger simulations is the fraction of major mergers (usually defined as those having a primary-to-secondary stellar mass ratio $\leq 3$), which is $\sim$50\% in the observations. More simulation work is needed to study the triggering of AGN activity in minor mergers \citep[e.g.,][]{YangC2019a}. 
 
\section{Summary} \label{sec:summary}

We have presented the first systematic X-ray study of optically selected AGN pairs in a statistical sample ($>10^2$) at redshift $z\sim0.1$. We have performed X-ray matches and detections with the \chandra\ data archive for a parent sample of \OIII -selected AGN pairs, a comparison sample of star-forming galaxy pairs, and a carefully drawn control sample of single AGNs in isolated galaxies. We summarize our main findings as follows:

\begin{enumerate}

\item We have detected at least one compact X-ray nuclei in 57 out of the 67 optically selected AGN pairs. The X-ray detection of both nuclei has confirmed AGN pairs in 21 optically selected systems, including 3 new discoveries of dual AGNs (i.e., AGN pairs with projected separations $<$10 kpc). The detection rate of X-ray nuclei is significantly higher in AGN pairs than in the comparison sample of star-forming galaxy pairs. 

\item We have measured the X-ray 2--10 keV luminosity for the detected X-ray nuclei based on the observed X-ray counts assuming a canonical power-law spectrum. 
The X-ray 2--10 keV luminosity of most AGN pairs (ranging between $3.9\times 10^{39}$ -- $6.8\times 10^{43} \rm~erg~s^{-1}$) also substantially exceeds the empirically predicted luminosity by star formation. Our results suggest that a large fraction of the X-ray-detected nuclei are genuine AGNs, lending support to their optical identification. 

\item We have studied the X-ray 2--10 keV luminosity of optically selected AGN pairs as a function of projected pair separation ($r_p$). The X-ray 2--10 keV luminosity increases as $r_p$ decreases from $\gtrsim$50 kpc to $\sim$15 kpc. However, it decreases with further decreasing $r_p$ at $r_p{<}$15 kpc, in contrast to theoretical predictions from merger simulations and to the empirical trend previously seen in \OIII\ luminosity. The apparent scarcity of X-ray luminous nuclei at $r_p$ smaller than a few kpc may be caused by heavy obscuration induced by merge-driven gas inflows. Alternatively, this may result from a negative feedback from merger-driven AGNs that already happens at the first pericentric passage and suppresses further black hole accretion or small number statistics.  

\end{enumerate}

Our results contribute to the understanding of the accretion and demographics of SMBHs in merging galaxies and the effects of merger-induced AGN on galaxy evolution. The X-ray confirmation of a statistical sample of dual AGNs is only enabled by the synergy between \chandra\ and SDSS, constituting an indispensable legacy for further more detailed studies. A continued effort of identifying such systems in larger numbers and characterizing their fundamental properties systematically in the X-rays will be highly rewarding.

\section*{Acknowledgements}

M.H. and Z.L. acknowledge support by the National Key Research and Development Program of China (2017YFA0402703). 
X.L. acknowledges support by NASA through Chandra Award Number GO3-14104X issued by the Chandra X-ray Observatory Center, which is operated by the Smithsonian Astrophysical
Observatory for and on behalf of NASA under contract NAS 8-03060.
We thank Yanmei Chen, Paul Green, Ralph Kraft and Jianfeng Wu for useful discussions, and Yali Zhu for help with the {\it Chandra} data preparation at the early stage of this work. 

This research has made use of software provided by the \chandra\ X-ray Center in the application packages CIAO, ChIPS, and Sherpa.

Funding for the Sloan Digital Sky Survey IV has been provided by the Alfred P. Sloan Foundation, the U.S. Department of Energy Office of Science, and the Participating Institutions. SDSS-IV acknowledges support and resources from the Center for High-Performance Computing at the University of Utah. The SDSS web site is www.sdss.org.

SDSS-IV is managed by the Astrophysical Research Consortium for the Participating Institutions of the SDSS Collaboration including the Brazilian Participation Group, the Carnegie Institution for Science, Carnegie Mellon University, the Chilean Participation Group, the French Participation Group, Harvard-Smithsonian Center for Astrophysics, Instituto de Astrof\'isica de Canarias, The Johns Hopkins University, Kavli Institute for the Physics and Mathematics of the Universe (IPMU) / University of Tokyo, Lawrence Berkeley National Laboratory, Leibniz Institut f\"ur Astrophysik Potsdam (AIP),  Max-Planck-Institut f\"ur Astronomie (MPIA Heidelberg), Max-Planck-Institut f\"ur Astrophysik (MPA Garching), Max-Planck-Institut f\"ur Extraterrestrische Physik (MPE), National Astronomical Observatories of China, New Mexico State University, New York University, University of Notre Dame, Observat\'ario Nacional / MCTI, The Ohio State University, Pennsylvania State University, Shanghai Astronomical Observatory, United Kingdom Participation Group,Universidad Nacional Aut\'onoma de M\'exico, University of Arizona, University of Colorado Boulder, University of Oxford, University of Portsmouth, University of Utah, University of Virginia, University of Washington, University of Wisconsin, Vanderbilt University, and Yale University.

Facilities: \chandra\ X-ray Observatory (ACIS), Sloan

\software{CIAO \citep{Fruscione2006}, Sherpa \citep{Freeman2001,Doe2007}, ChiPS \citep{Germain2006}}

\section*{Appendix}

In this appendix, we present three new dual AGN systems uncovered in our X-ray survey whose X-ray properties have not been reported in the literature. Below we discuss these systems in detail. The optical and X-ray images are shown in Figure \ref{fig:image_AGNpair}. Table \ref{table:newlylx} lists their physical measurements.

SDSS~J1450$-$0002 The separation of this AGN pair is 2.3 kpc as shown in Figure \ref{fig:image_AGNpair}. Both galaxies in the merger are optically classified as H\,{\tiny II}/AGN composites or LINERs. The SW nucleus (J1450-0002a) is detected in both the $S$ and $H$ bands, whereas the NE nucleus (J1450-0002b) is detected in the $F$ band only. For the SW nucleus, the comparison between the observed to the expected star-formation-induced X-ray luminosities (Table \ref{table:newlylx}) are consistent with the AGN scenario. For the NE nucleus, on the other hand, the upper limits of the X-ray luminosity in the $S$ and $H$ bands are still consistent with the presence of an additional AGN, although the possibility that one AGN in the NW nucleus ionizes gas in both galaxies cannot be ruled out.

SDSS~J0858$+$1822 This is a candidate triple AGN system. But the third nucleus is not included in SDSS DR7. The separation (3.3 kpc) of this system is calculated between nuclei A and B as shown in Figure \ref{fig:image_AGNpair}. The NE nucleus (J0858+1822b) in the merger is optically classified as a Type 2 Seyfert whereas the SW nucleus (J0858+1822a) a Type 2 Seyfert or a LINER. Both NE and SW nuclei are detected in the $S$ band only. However, the X-ray counterpart of SW nucleus looks more likely correlated to a third nucleus between pairs. For both nuclei, the observed X-ray luminosities are similar to those expected from star formation-induced X-ray luminosities (Table \ref{table:newlylx}), although an additional AGN component cannot be ruled out given significant uncertainties in the estimate.


SDSS~J1414$-$0000 The separation of this AGN pair is 4.9 kpc as shown in Figure \ref{fig:image_AGNpair}. The SW nucleus (J1414-0000a) in the merger is optically classified as a Type 2 Seyfert whereas the NE nucleus (J1414-0000b) a H\,{\tiny II}/AGN composite. Both nuclei are detected in both the $S$ and $H$ bands and are luminous. For both nuclei, the expected star-formation-induced X-ray luminosities are too low to explain the observed values in both the $S$ and $H$ bands (Table \ref{table:newlylx}), consistent with the AGN scenario.

\bibliography{binaryrefs}

\startlongtable
\begin{deluxetable}{cccccccccccc}
\tabletypesize{\scriptsize}
\tablecaption{Information of AGN pairs with {\it Chandra} observation \label{tab:info}}
\tablewidth{0pt}
\tablehead{
\colhead{Name} & \colhead{R.A.} & \colhead{Dec.} & \colhead{$z$} & \colhead{$r_{\rm p}$} & \colhead{log$M_{\ast}$} & \colhead{$F_{\rm \OIII}$} &  \colhead{SFR} & \colhead{ObsID} & \colhead{Exp.} & \colhead{log $L_{\rm X,lim}$} & \colhead{flag}
}
\colnumbers
\startdata
 J145059.71-000215.2  & 222.74879  &  -0.03756  & 0.0429  &  2.3  & 11.4  & $ 103.2 \pm  7.8$  &  $ 0.11^{+ 0.45}_{- 0.11}$  & 14008 &  25.1  &  40.26  & 1 \\
 J145059.84-000213.2  & 222.74933  &  -0.03700  & 0.0430  &  2.3  &  9.7  & $  29.1 \pm  6.7$  &  $ 0.02^{+ 0.10}_{- 0.02}$  & 14008 &  25.1  &  40.27  & 1 \\
 J102700.40+174901.0  & 156.75167  &  17.81694  & 0.0665  &  3.0  & 10.9  & $  72.3 \pm  5.9$  &  $ 1.09^{+ 3.04}_{- 0.97}$  & 14971 &  49.1  &  40.35  & 0 \\
 J102700.56+174900.3  & 156.75233  &  17.81675  & 0.0666  &  3.0  &  ...  & $  95.1 \pm  2.5$  &  $12.86^{+29.44}_{- 9.38}$  & 14971 &  49.1  &  40.35  & 1 \\
 J085837.53+182221.6  & 134.65637  &  18.37267  & 0.0587  &  3.3  & 10.4  & $ 399.7 \pm  6.3$  &  $ 3.55^{+ 8.31}_{- 2.62}$  & 14970 &  19.8  &  40.49  & 1 \\
 J085837.68+182223.4  & 134.65700  &  18.37317  & 0.0589  &  3.3  & 11.1  & $1096.4 \pm 11.7$  &  $ 3.46^{+ 9.05}_{- 2.88}$  & 14970 &  19.8  &  40.49  & 1 \\
 J154403.45+044607.5  & 236.01437  &   4.76875  & 0.0420  &  3.4  &  9.8  & $ 137.4 \pm  5.3$  &  $ 0.37^{+ 1.09}_{- 0.34}$  & 14968 &  14.9  &  40.32  & 1 \\
 J154403.67+044610.1  & 236.01529  &   4.76947  & 0.0416  &  3.4  & 11.1  & $ 232.2 \pm  4.6$  &  $ 0.91^{+ 2.27}_{- 0.72}$  & 14968 &  14.9  &  40.31  & 1 \\
 J105842.44+314457.6  & 164.67683  &  31.74933  & 0.0728  &  4.1  & 10.0  & $ 225.0 \pm  5.2$  &  $ 1.95^{+ 2.73}_{- 1.19}$  & 14966 &  14.5  &  40.82  & 1 \\
 J105842.58+314459.8  & 164.67742  &  31.74994  & 0.0723  &  4.1  & 10.9  & $ 619.8 \pm  8.4$  &  $ 2.44^{+ 5.33}_{- 1.75}$  & 14966 &  14.5  &  40.81  & 1 \\
 J002208.69+002200.5  &   5.53621  &   0.36681  & 0.0710  &  4.2  & 11.0  & $  25.9 \pm  4.3$  &  $ 0.02^{+ 0.15}_{- 0.02}$  &  2252 &  70.9  &  40.43  & 1 \\
 J002208.83+002202.8  &   5.53679  &   0.36744  & 0.0707  &  4.2  & 11.2  & $  18.5 \pm  3.2$  &  $ 0.03^{+ 0.16}_{- 0.02}$  &  2252 &  70.9  &  40.43  & 1 \\
 J220634.97+000327.6  & 331.64571  &   0.05767  & 0.0466  &  4.3  &  ...  & $  85.3 \pm  5.9$  &  $ 0.68^{+ 1.58}_{- 0.50}$  & 19454 &  15.9  &  40.47  & 0 \\
 J220635.08+000323.2  & 331.64617  &   0.05644  & 0.0461  &  4.3  &  ...  & $ 290.4 \pm  6.9$  &  $ 0.74^{+ 1.86}_{- 0.59}$  & 19454 &  15.9  &  40.46  & 1 \\
 J133031.75-003611.9  & 202.63229  &  -0.60331  & 0.0542  &  4.4  &  8.8  & $ 233.0 \pm  4.3$  &  $ 0.35^{+ 0.49}_{- 0.21}$  & 14967 &  14.9  &  40.54  & 0 \\
 J133032.00-003613.5  & 202.63333  &  -0.60375  & 0.0542  &  4.4  & 10.7  & $ 815.4 \pm  9.7$  &  $ 4.58^{+ 7.86}_{- 2.98}$  & 14967 &  14.9  &  40.54  & 1 \\
 J121345.95+024839.0  & 183.44146  &   2.81083  & 0.0730  &  4.8  & 10.4  & $  42.6 \pm  2.6$  &  $ 9.29^{+19.88}_{- 6.59}$  &  4110 &  10.0  &  40.86  & 1 \\
 J121346.11+024841.4  & 183.44212  &   2.81150  & 0.0731  &  4.8  & 10.6  & $ 230.8 \pm  4.1$  &  $13.28^{+17.60}_{- 7.80}$  &  4110 &  10.0  &  40.86  & 1 \\
 J141447.15-000013.3  & 213.69646  &  -0.00369  & 0.0475  &  4.9  & 10.5  & $ 647.2 \pm  7.1$  &  $ 0.27^{+ 0.77}_{- 0.24}$  & 17149 &  67.0  &  40.03  & 1 \\
 J141447.48-000011.3  & 213.69783  &  -0.00314  & 0.0474  &  4.9  & 10.2  & $ 148.2 \pm  4.3$  &  $ 3.54^{+ 4.94}_{- 2.12}$  & 17149 &  67.0  &  40.02  & 1 \\
 J080523.29+281815.8  & 121.34704  &  28.30439  & 0.1284  &  5.6  & 11.2  & $2910.7 \pm 18.4$  &  $11.68^{+15.90}_{- 6.95}$  & 14964 &  14.0  &  41.35  & 1 \\
 J080523.40+281814.1  & 121.34750  &  28.30392  & 0.1286  &  5.6  &  9.8  & $ 257.3 \pm  6.5$  &  $ 2.24^{+ 3.31}_{- 1.38}$  & 14964 &  14.0  &  41.35  & 0 \\
 J133817.27+481632.3  & 204.57196  &  48.27564  & 0.0278  &  6.4  & 10.0  & $4901.8 \pm 34.8$  &  $ 5.48^{+ 7.89}_{- 3.33}$  &  2044 &  10.9  &  40.06  & 1 \\
 J133817.77+481641.1  & 204.57404  &  48.27808  & 0.0277  &  6.4  & 10.6  & $4142.2 \pm 31.6$  &  $ 2.84^{+ 3.66}_{- 1.66}$  &  2044 &  10.9  &  40.05  & 1 \\
 J114753.63+094552.0  & 176.97346  &   9.76444  & 0.0951  &  6.6  & 10.3  & $1368.1 \pm 13.0$  &  $ 8.63^{+14.39}_{- 5.58}$  & 18198 &  22.6  &  41.00  & 1 \\
 J114753.68+094555.6  & 176.97367  &   9.76544  & 0.0966  &  6.6  & 11.0  & $  35.7 \pm  5.8$  &  $ 1.10^{+ 1.56}_{- 0.63}$  & 18198 &  22.6  &  41.01  & 0 \\
 J005113.94+002047.2  &  12.80808  &   0.34644  & 0.1124  &  6.9  &  ...  & $  55.4 \pm  3.0$  &  $ 3.19^{+ 6.38}_{- 2.21}$  & 19453 &  28.7  &  41.01  & 1 \\
 J005114.12+002049.2  &  12.80883  &   0.34700  & 0.1126  &  6.9  & 11.0  & $  31.5 \pm  1.5$  &  $ 8.23^{+24.20}_{- 6.63}$  & 19453 &  28.7  &  41.01  & 1 \\
 J085312.36+162619.5  & 133.30150  &  16.43875  & 0.0637  &  8.0  & 10.5  & $ 278.0 \pm  6.3$  &  $ 3.57^{+ 4.88}_{- 2.12}$  & 12143 &   9.9  &  40.78  & 1 \\
 J085312.70+162615.5  & 133.30292  &  16.43764  & 0.0649  &  8.0  &  9.9  & $ 362.6 \pm  3.9$  &  $ 0.32^{+ 0.49}_{- 0.20}$  & 12143 &   9.9  &  40.80  & 1 \\
 J083847.57+040734.6  & 129.69821  &   4.12628  & 0.0484  &  8.1  &  9.9  & $  27.7 \pm  3.3$  &  $ 0.24^{+ 0.58}_{- 0.18}$  & 18159 &  14.8  &  40.50  & 0 \\
 J083848.15+040734.1  & 129.70062  &   4.12614  & 0.0476  &  8.1  & 10.8  & $2413.7 \pm 23.7$  &  $ 0.66^{+ 1.67}_{- 0.53}$  & 18159 &  14.8  &  40.49  & 1 \\
 J083817.59+305453.5  & 129.57329  &  30.91486  & 0.0478  &  8.8  & 10.7  & $  80.2 \pm  4.5$  &  $ 2.62^{+ 5.73}_{- 1.88}$  & 12817 &   4.9  &  40.93  & 1 \\
 J083817.95+305501.1  & 129.57479  &  30.91697  & 0.0481  &  8.8  & 11.2  & $  20.9 \pm  4.9$  &  $ 0.05^{+ 0.28}_{- 0.04}$  & 12817 &   4.9  &  40.94  & 0 \\
 J090714.45+520343.4  & 136.81021  &  52.06206  & 0.0596  &  8.9  & 10.6  & $ 436.2 \pm  6.2$  &  $ 0.59^{+ 1.61}_{- 0.52}$  & 14965 &  14.5  &  40.64  & 1 \\
 J090714.61+520350.7  & 136.81087  &  52.06408  & 0.0602  &  8.9  & 10.3  & $ 325.1 \pm  3.4$  &  $ 1.28^{+ 2.95}_{- 0.94}$  & 14965 &  14.5  &  40.65  & 1 \\
 J134736.41+173404.7  & 206.90171  &  17.56797  & 0.0447  &  9.3  &  9.4  & $1681.9 \pm 13.7$  &  $ 0.30^{+ 0.41}_{- 0.18}$  & 10561 &   1.7  &  41.17  & 1 \\
 J134737.11+173404.1  & 206.90462  &  17.56781  & 0.0450  &  9.3  & 10.5  & $  23.1 \pm  2.8$  &  $ 0.13^{+ 0.35}_{- 0.11}$  & 10561 &   1.7  &  41.17  & 0 \\
 J000249.07+004504.8  &   0.70446  &   0.75133  & 0.0868  &  9.5  & 11.2  & $1037.7 \pm 11.6$  &  $ 0.16^{+ 0.97}_{- 0.15}$  &  4861 &   5.0  &  41.43  & 1 \\
 J000249.44+004506.7  &   0.70600  &   0.75186  & 0.0865  &  9.5  & 10.9  & $  17.1 \pm  3.9$  &  $ 0.01^{+ 0.09}_{- 0.01}$  &  4861 &   5.0  &  41.43  & 0 \\
 J135429.06+132757.3  & 208.62108  &  13.46592  & 0.0633  & 12.5  & 10.1  & $3422.6 \pm 25.5$  &  $ 0.04^{+ 0.12}_{- 0.04}$  & 16115 &   9.1  &  40.86  & 1 \\
 J135429.18+132807.4  & 208.62158  &  13.46872  & 0.0634  & 12.5  & 10.7  & $ 187.2 \pm  4.1$  &  $ 0.64^{+ 1.66}_{- 0.52}$  & 16115 &   9.1  &  40.86  & 0 \\
 J135225.64+142919.3  & 208.10683  &  14.48869  & 0.0415  & 14.1  & 10.7  & $  45.6 \pm  5.7$  &  $ 0.01^{+ 0.07}_{- 0.01}$  & 12822 &   4.9  &  40.86  & 0 \\
 J135226.65+142927.5  & 208.11104  &  14.49097  & 0.0406  & 14.1  & 11.2  & $  43.4 \pm  5.4$  &  $ 3.74^{+ 7.98}_{- 2.65}$  & 12822 &   4.9  &  40.84  & 0 \\
 J124610.11+304354.9  & 191.54212  &  30.73192  & 0.0219  & 16.4  & 10.9  & $  82.2 \pm  3.4$  &  $10.61^{+19.49}_{- 7.13}$  &  2043 &  27.9  &  39.38  & 1 \\
 J124611.24+304321.9  & 191.54683  &  30.72275  & 0.0218  & 16.4  &  9.9  & $ 206.3 \pm  6.9$  &  $ 0.01^{+ 0.05}_{- 0.01}$  &  2043 &  27.9  &  39.38  & 1 \\
 J111519.23+542310.9  & 168.83012  &  54.38636  & 0.0713  & 17.1  & 10.5  & $ 345.5 \pm  4.4$  &  $ 0.01^{+ 0.04}_{- 0.01}$  & 13902 &  19.1  &  40.73  & 0 \\
 J111519.98+542316.7  & 168.83325  &  54.38797  & 0.0704  & 17.1  & 11.1  & $1054.1 \pm 10.7$  &  $ 1.72^{+ 5.20}_{- 1.60}$  & 13902 &  19.1  &  40.72  & 1 \\
 J143454.22+334934.5  & 218.72592  &  33.82625  & 0.0578  & 18.0  & 10.8  & $ 127.9 \pm  4.5$  &  $ 0.01^{+ 0.07}_{- 0.01}$  & 19648 &  24.7  &  40.92  & 0 \\
 J143454.68+334920.0  & 218.72783  &  33.82222  & 0.0587  & 18.0  & 10.7  & $  16.9 \pm  3.7$  &  $ 0.64^{+ 1.82}_{- 0.58}$  & 19648 &  24.7  &  40.93  & 0 \\
 J111830.28+402554.0  & 169.62617  &  40.43167  & 0.1545  & 18.1  &  ...  &     ...            &  $-1.00^{+ 0.00}_{- 0.00}$  &   868 &  16.5  &  41.47  & 1 \\
 J111830.68+402557.0  & 169.62783  &  40.43250  & 0.1559  & 18.1  & 11.0  & $  79.5 \pm  3.5$  &  $ 8.53^{+22.78}_{- 6.52}$  &   868 &  16.5  &  41.48  & 0 \\
 J214622.41+000452.1  & 326.59337  &   0.08114  & 0.0754  & 18.7  & 10.4  & $  16.8 \pm  2.4$  &  $ 4.72^{+10.33}_{- 3.39}$  & 20466 &  11.9  &  41.20  & 0 \\
 J214623.23+000456.7  & 326.59679  &   0.08242  & 0.0750  & 18.7  & 10.9  & $ 109.6 \pm  3.6$  &  $ 2.72^{+ 6.49}_{- 2.04}$  & 20466 &  11.9  &  41.19  & 1 \\
 J085441.74+305754.7  & 133.67392  &  30.96519  & 0.1970  & 20.0  &  ...  &     ...            &  $-1.00^{+ 0.00}_{- 0.00}$  & 11662 &  10.4  &  42.05  & 1 \\
 J085442.11+305757.4  & 133.67546  &  30.96594  & 0.1956  & 20.0  & 10.9  & $  50.2 \pm  2.9$  &  $ 1.89^{+ 4.50}_{- 1.40}$  & 11662 &  10.4  &  42.04  & 0 \\
 J142947.66+353427.5  & 217.44858  &  35.57431  & 0.0283  & 25.0  & 11.0  & $ 246.9 \pm 11.8$  &  $ 0.01^{+ 0.07}_{- 0.01}$  &  7776 &  13.9  &  40.15  & 1 \\
 J142950.32+353412.2  & 217.45967  &  35.57006  & 0.0292  & 25.0  & 10.5  & $  26.2 \pm  4.3$  &  $ 0.12^{+ 0.50}_{- 0.12}$  &  7776 &  13.9  &  40.18  & 0 \\
 J155548.02+242911.9  & 238.95008  &  24.48664  & 0.0705  & 28.8  & 10.5  & $  13.5 \pm  2.9$  &  $ 0.28^{+ 0.35}_{- 0.14}$  &  3984 &  20.0  &  40.72  & 0 \\
 J155549.45+242911.5  & 238.95604  &  24.48653  & 0.0705  & 28.8  & 10.5  & $ 253.8 \pm  6.3$  &  $ 0.39^{+ 1.08}_{- 0.35}$  &  3984 &  20.0  &  40.72  & 1 \\
 J001842.19-101613.9  &   4.67579  & -10.27053  & 0.1215  & 32.0  & 10.9  & $  17.9 \pm  2.6$  &  $ 1.04^{+ 2.84}_{- 0.91}$  & 13823 &  29.6  &  41.27  & 0 \\
 J001842.81-101625.1  &   4.67837  & -10.27364  & 0.1220  & 32.0  & 10.9  & $  15.9 \pm  3.7$  &  $ 0.50^{+ 1.39}_{- 0.44}$  & 13823 &  29.6  &  41.27  & 0 \\
 J160527.62+174858.4  & 241.36508  &  17.81622  & 0.0335  & 32.4  &  9.7  & $  18.4 \pm  2.2$  &  $ 0.01^{+ 0.01}_{- 0.00}$  & 17038 &   7.7  &  40.38  & 0 \\
 J160527.91+174946.7  & 241.36629  &  17.82964  & 0.0337  & 32.4  & 10.1  & $  37.1 \pm  2.9$  &  $ 0.11^{+ 0.31}_{- 0.10}$  & 17038 &   7.7  &  40.39  & 0 \\
 J125929.96+275723.2  & 194.87483  &  27.95644  & 0.0227  & 34.4  & 10.6  & $  33.3 \pm  8.6$  &  $ 0.01^{+ 0.03}_{- 0.00}$  &  13996* &  592.2  &   39.40  &  1 \\
 J125934.12+275648.6  & 194.89217  &  27.94683  & 0.0240  & 34.4  & 10.6  & $  43.8 \pm 10.6$  &  $ 0.08^{+ 0.10}_{- 0.04}$  &  13996* & 592.2  &  39.45  & 1 \\
 J131513.87+442426.5  & 198.80779  &  44.40736  & 0.0354  & 35.9  & 10.5  & $ 414.5 \pm  8.3$  &  $ 0.45^{+ 1.10}_{- 0.35}$  & 12242 &  29.6  &  39.89  & 1 \\
 J131517.27+442425.6  & 198.82196  &  44.40711  & 0.0355  & 35.9  & 10.9  & $2250.4 \pm 21.5$  &  $ 0.34^{+ 1.27}_{- 0.33}$  & 12242 &  29.6  &  39.89  & 1 \\
 J142626.63+330439.5  & 216.61096  &  33.07764  & 0.0636  & 37.1  & 10.4  & $  13.6 \pm  2.8$  &  $17.39^{+44.72}_{-15.12}$  & 19663 &  24.0  &  40.67  & 0 \\
 J142627.54+330412.8  & 216.61475  &  33.07022  & 0.0645  & 37.1  & 10.4  & $  60.9 \pm  3.0$  &  $ 0.03^{+ 0.12}_{- 0.03}$  & 19663 &  24.0  &  40.68  & 1 \\
 J011544.85+001400.0  &  18.93687  &   0.23333  & 0.0445  & 39.4  & 10.6  & $ 252.0 \pm  5.6$  &  $ 0.33^{+ 0.91}_{- 0.29}$  &  3204 &  37.6  &  40.38  & 1 \\
 J011545.23+001444.6  &  18.93846  &   0.24572  & 0.0431  & 39.4  &  ...  & $   7.6 \pm  2.2$  &  $ 0.01^{+ 0.06}_{- 0.01}$  &  3204 &  37.6  &  40.35  & 0 \\
 J112545.05+144035.7  & 171.43771  &  14.67658  & 0.0340  & 49.7  & 11.0  & $ 108.1 \pm  4.7$  &  $17.15^{+27.03}_{-10.85}$  & 15053 &  14.5  &  40.10  & 1 \\
 J112549.55+144006.6  & 171.45646  &  14.66850  & 0.0339  & 49.7  & 10.9  & $  60.4 \pm  3.0$  &  $12.46^{+23.30}_{- 8.40}$  & 15053 &  14.5  &  40.09  & 1 \\
 J122726.27+081123.8  & 186.85946  &   8.18994  & 0.1185  & 50.0  &  ...  & $  10.0 \pm  2.4$  &  $ 2.39^{+ 3.12}_{- 1.30}$  &  8060 &   5.1  &  41.64  & 0 \\
 J122727.54+081137.0  & 186.86475  &   8.19361  & 0.1194  & 50.0  &  ...  & $  10.9 \pm  2.4$  &  $ 0.90^{+ 1.19}_{- 0.49}$  &  8060 &   5.1  &  41.65  & 0 \\
 J145806.24+013020.5  & 224.52600  &   1.50569  & 0.0429  & 51.3  & 11.1  & $  92.9 \pm  8.2$  &  $ 0.01^{+ 0.09}_{- 0.01}$  & 11325 &   8.3  &  40.68  & 0 \\
 J145808.49+013110.9  & 224.53537  &   1.51969  & 0.0444  & 51.3  & 10.8  & $  27.0 \pm  3.5$  &  $ 0.01^{+ 0.09}_{- 0.01}$  & 11325 &   8.3  &  40.71  & 0 \\
 J143310.55+525830.5  & 218.29396  &  52.97514  & 0.0476  & 52.3  &  ...  &     ...            &  $-1.00^{+ 0.00}_{- 0.00}$  & 11473 &   1.8  &  41.18  & 1 \\
 J143312.96+525747.3  & 218.30400  &  52.96314  & 0.0473  & 52.3  & 10.9  & $  93.3 \pm  6.9$  &  $ 0.31^{+ 0.90}_{- 0.28}$  & 11473 &   1.8  &  41.17  & 1 \\
 J032015.52+412355.1  &  50.06467  &  41.39864  & 0.0224  & 58.0  & 11.0  & $  82.6 \pm 10.2$  &  $ 0.01^{+ 0.05}_{- 0.01}$  & 12037* & 197.3  &  39.83  & 1 \\
 J032020.97+412225.5  &  50.08737  &  41.37375  & 0.0237  & 58.0  &  9.8  & $  51.2 \pm  8.2$  &  $ 0.00^{+ 0.03}_{- 0.00}$  & 12037* & 197.3  &  39.88  & 0 \\
 J032512.84+003711.6  &  51.30350  &   0.61989  & 0.1502  & 59.7  & 11.1  & $  16.8 \pm  3.7$  &  $ 0.03^{+ 0.21}_{- 0.03}$  &  7733 &   4.5  &  41.91  & 1 \\
 J032513.30+003733.3  &  51.30542  &   0.62592  & 0.1483  & 59.7  & 10.8  & $  55.5 \pm  3.7$  &  $ 0.11^{+ 0.46}_{- 0.10}$  &  7733 &   4.5  &  41.90  & 0 \\
 J135740.18+223137.7  & 209.41742  &  22.52714  & 0.0622  & 60.8  & 10.8  & $  36.2 \pm  3.5$  &  $ 0.03^{+ 0.22}_{- 0.03}$  & 20115 &  10.0  &  40.88  & 1 \\
 J135742.12+223056.1  & 209.42550  &  22.51558  & 0.0621  & 60.8  & 10.8  & $ 188.3 \pm  4.3$  &  $ 0.53^{+ 1.60}_{- 0.50}$  & 20115 &  10.0  &  40.88  & 0 \\
 J152131.56+074425.8  & 230.38150  &   7.74050  & 0.0425  & 63.0  & 10.9  & $  19.1 \pm  5.1$  &  $ 0.02^{+ 0.10}_{- 0.02}$  &   900 &  57.3  &  40.38  & 1 \\
 J152136.39+074420.7  & 230.40162  &   7.73908  & 0.0441  & 63.0  & 10.6  & $  13.4 \pm  4.2$  &  $ 0.01^{+ 0.08}_{- 0.01}$  &   900 &  57.3  &  40.41  & 0 \\
 J155957.11+412840.0  & 239.98796  &  41.47778  & 0.0333  & 64.1  & 10.8  & $  32.3 \pm  4.7$  &  $ 0.01^{+ 0.05}_{- 0.01}$  &  9206 &  10.0  &  40.24  & 0 \\
 J160003.55+412845.4  & 240.01479  &  41.47928  & 0.0335  & 64.1  & 10.9  & $ 114.4 \pm  9.7$  &  $ 0.03^{+ 0.20}_{- 0.03}$  &  9206 &  10.0  &  40.25  & 1 \\
 J092458.83+022519.7  & 141.24512  &   2.42214  & 0.0716  & 64.3  & 10.8  & $  45.6 \pm  3.3$  &  $ 1.13^{+ 2.12}_{- 0.77}$  & 11564* & 124.2  &  40.31  & 1 \\
 J092501.12+022447.5  & 141.25467  &   2.41319  & 0.0716  & 64.3  & 10.4  & $   7.7 \pm  1.9$  &  $ 0.25^{+ 0.31}_{- 0.13}$  & 11564* & 124.2  &  40.31  & 0 \\
 J160526.61+175435.6  & 241.36087  &  17.90989  & 0.0410  & 64.5  & 10.9  & $  41.9 \pm  4.6$  &  $ 0.01^{+ 0.05}_{- 0.01}$  & 17038 &   7.7  &  40.56  & 0 \\
 J160529.43+175543.0  & 241.37262  &  17.92861  & 0.0400  & 64.5  & 10.9  & $  42.7 \pm  4.8$  &  $ 0.16^{+ 0.65}_{- 0.15}$  & 17038 &   7.7  &  40.54  & 0 \\
 J150457.12+260058.5  & 226.23800  &  26.01625  & 0.0540  & 65.2  & 11.3  & $ 119.4 \pm  5.4$  &  $ 0.01^{+ 0.07}_{- 0.01}$  & 11845 &  56.9  &  40.17  & 1 \\
 J150501.22+260101.5  & 226.25508  &  26.01708  & 0.0545  & 65.2  & 10.9  & $  10.8 \pm  3.3$  &  $ 0.17^{+ 0.23}_{- 0.10}$  & 11845 &  56.9  &  40.18  & 1 \\
 J123405.05+062933.2  & 188.52104  &   6.49256  & 0.0809  & 65.9  &  ...  & $  22.3 \pm  4.4$  &  $ 0.05^{+ 0.24}_{- 0.05}$  & 19407 &   8.0  &  41.31  & 0 \\
 J123407.84+062922.5  & 188.53267  &   6.48958  & 0.0808  & 65.9  &  ...  & $  24.6 \pm  3.5$  &  $ 0.03^{+ 0.16}_{- 0.02}$  & 19407 &   8.0  &  41.31  & 1 \\
 J151018.41+074219.6  & 227.57671  &   7.70544  & 0.0424  & 67.3  & 11.3  & $ 104.0 \pm  7.1$  &  $ 0.09^{+ 0.52}_{- 0.08}$  &   939 &   4.9  &  40.79  & 0 \\
 J151022.87+074304.2  & 227.59529  &   7.71783  & 0.0442  & 67.3  & 11.0  & $  85.2 \pm  9.6$  &  $ 0.02^{+ 0.12}_{- 0.02}$  &   939 &   4.9  &  40.82  & 1 \\
 J211702.44-064507.3  & 319.26017  &  -6.75203  & 0.0298  & 71.9  & 10.7  & $  24.1 \pm  5.7$  &  $ 0.15^{+ 0.39}_{- 0.13}$  & 20435 &  12.9  &  40.13  & 0 \\
 J211706.99-064327.9  & 319.27912  &  -6.72442  & 0.0292  & 71.9  & 11.1  & $ 274.2 \pm 15.8$  &  $ 0.98^{+ 2.71}_{- 0.87}$  & 20435 &  12.9  &  40.11  & 1 \\
 J140343.63+292045.3  & 210.93179  &  29.34592  & 0.0633  & 76.3  & 11.2  & $  54.5 \pm  3.9$  &  $ 0.40^{+ 2.19}_{- 0.38}$  &  4782 &   2.6  &  41.26  & 1 \\
 J140345.02+292144.0  & 210.93758  &  29.36222  & 0.0637  & 76.3  & 11.1  & $ 618.3 \pm  9.0$  &  $ 3.01^{+ 7.72}_{- 2.45}$  &  4782 &   2.6  &  41.26  & 1 \\
 J121351.33+024805.8  & 183.46387  &   2.80161  & 0.0747  & 76.9  & 10.8  & $  15.8 \pm  2.9$  &  $ 0.11^{+ 0.40}_{- 0.11}$  &  4110 &  10.0  &  40.88  & 0 \\
 J121354.87+024753.0  & 183.47862  &   2.79806  & 0.0743  & 76.9  & 10.7  & $ 711.8 \pm  9.0$  &  $ 7.65^{+10.99}_{- 4.63}$  &  4110 &  10.0  &  40.87  & 1 \\
 J032224.28+421721.7  &  50.60117  &  42.28936  & 0.0510  & 77.0  & 10.9  & $  38.8 \pm  3.0$  &  $ 0.90^{+ 2.29}_{- 0.73}$  & 17273 &   5.0  &  41.08  & 0 \\
 J032227.88+421626.4  &  50.61617  &  42.27400  & 0.0516  & 77.0  &  9.9  & $ 928.1 \pm  7.0$  &  $ 2.16^{+ 4.12}_{- 1.47}$  & 17273 &   5.0  &  41.09  & 1 \\
 J160501.37+174632.5  & 241.25571  &  17.77569  & 0.0330  & 78.1  & 10.9  & $  54.8 \pm  8.0$  &  $ 0.01^{+ 0.06}_{- 0.01}$  & 20087* &  82.9  &  40.04  & 1 \\
 J160507.89+174527.6  & 241.28287  &  17.75767  & 0.0334  & 78.1  & 11.3  & $  52.8 \pm  5.0$  &  $ 0.05^{+ 0.31}_{- 0.05}$  & 20087* &  82.9  &  40.05  & 0 \\
 J093325.71+340253.1  & 143.35712  &  34.04808  & 0.0272  & 78.6  & 11.0  & $  51.0 \pm  6.6$  &  $ 0.01^{+ 0.04}_{- 0.01}$  &  3227 &  34.3  &  39.70  & 1 \\
 J093334.42+340153.4  & 143.39342  &  34.03150  & 0.0287  & 78.6  &  9.9  & $   5.4 \pm  1.8$  &  $ 0.01^{+ 0.01}_{- 0.01}$  &  3227 &  34.3  &  39.75  & 0 \\
 J104229.80+050500.2  & 160.62417  &   5.08339  & 0.0281  & 80.3  &  9.6  & $   7.9 \pm  1.9$  &  $ 0.01^{+ 0.00}_{- 0.00}$  & 13871 &  49.0  &  39.58  & 0 \\
 J104232.05+050241.9  & 160.63354  &   5.04497  & 0.0272  & 80.3  &  9.2  & $ 459.3 \pm  5.9$  &  $ 0.06^{+ 0.15}_{- 0.05}$  & 13871 &  49.0  &  39.55  & 1 \\
 J091334.02+300223.4  & 138.39175  &  30.03983  & 0.0214  & 81.2  &  9.9  & $  10.0 \pm  2.5$  &  $ 0.01^{+ 0.01}_{- 0.01}$  &  5789 &  17.9  &  39.52  & 0 \\
 J091339.47+295934.7  & 138.41446  &  29.99297  & 0.0225  & 81.2  & 11.2  & $ 300.6 \pm 19.3$  &  $ 0.01^{+ 0.04}_{- 0.01}$  &  5789 &  17.9  &  39.57  & 1 \\
 J004239.54-093053.8  &  10.66475  &  -9.51494  & 0.0574  & 85.7  & 10.8  & $  18.8 \pm  4.3$  &  $ 0.19^{+ 0.25}_{- 0.10}$  &  4881 &   9.8  &  41.01  & 0 \\
 J004244.64-093102.8  &  10.68600  &  -9.51744  & 0.0568  & 85.7  & 10.7  & $  15.7 \pm  4.4$  &  $ 0.02^{+ 0.10}_{- 0.02}$  &  4881 &   9.8  &  41.00  & 0 \\
 J215945.57-002712.0  & 329.93987  &  -0.45333  & 0.1273  & 86.6  & 10.9  & $  19.9 \pm  2.8$  &  $ 0.11^{+ 0.42}_{- 0.11}$  & 11509 &   7.8  &  41.73  & 1 \\
 J215947.31-002644.3  & 329.94712  &  -0.44564  & 0.1281  & 86.6  &  ...  & $1072.6 \pm  9.5$  &  $ 1.35^{+ 4.34}_{- 1.27}$  & 11509 &   7.8  &  41.74  & 1 \\
 J015303.50+010123.0  &  28.26458  &   1.02306  & 0.0595  & 88.5  & 10.6  & $  15.9 \pm  2.6$  &  $ 0.04^{+ 0.17}_{- 0.04}$  &  1448 &   6.9  &  40.96  & 0 \\
 J015308.29+010150.7  &  28.28454  &   1.03075  & 0.0602  & 88.5  & 10.9  & $ 109.5 \pm  4.2$  &  $ 0.02^{+ 0.10}_{- 0.01}$  &  1448 &   6.9  &  40.97  & 0 \\
 J082321.67+042220.9  & 125.84029  &   4.37247  & 0.0311  & 88.7  & 11.4  & $  38.3 \pm  8.4$  &  $ 0.01^{+ 0.04}_{- 0.01}$  & 15161 &   9.9  &  40.54  & 1 \\
 J082329.88+042333.0  & 125.87450  &   4.39250  & 0.0308  & 88.7  & 10.3  & $ 157.0 \pm  5.8$  &  $ 0.04^{+ 0.20}_{- 0.04}$  & 15161 &   9.9  &  40.53  & 0 \\
 J080529.88+241004.4  & 121.37450  &  24.16789  & 0.0595  & 89.8  & 11.0  & $  46.8 \pm  5.5$  &  $ 0.01^{+ 0.09}_{- 0.01}$  & 19496* & 177.3  &  39.85  & 1 \\
 J080535.00+240950.3  & 121.39583  &  24.16397  & 0.0597  & 89.8  & 11.0  & $2791.1 \pm 23.2$  &  $ 0.02^{+ 0.10}_{- 0.01}$  & 19496* & 177.3  &  39.86  & 1 \\
 J135408.85+325434.6  & 208.53687  &  32.90961  & 0.0256  & 95.4  &  9.1  & $  15.3 \pm  1.2$  &  $ 0.01^{+ 0.00}_{- 0.00}$  &  3004 &   7.0  &  40.09  & 0 \\
 J135419.95+325547.8  & 208.58312  &  32.92994  & 0.0261  & 95.4  & 10.9  & $1139.1 \pm 17.2$  &  $ 0.50^{+ 1.38}_{- 0.45}$  &  3004 &   7.0  &  40.10  & 1 \\
 J102141.89+130550.4  & 155.42454  &  13.09733  & 0.0765  & 97.2  & 10.7  & $ 148.8 \pm  3.3$  &  $ 6.29^{+15.73}_{- 4.83}$  &  4107 &   5.4  &  41.10  & 0 \\
 J102142.79+130656.1  & 155.42829  &  13.11558  & 0.0763  & 97.2  & 11.1  & $  25.4 \pm  3.1$  &  $14.83^{+35.11}_{-11.04}$  &  4107 &   5.4  &  41.09  & 0 \\
 J125939.94+312707.4  & 194.91642  &  31.45206  & 0.0514  & 97.3  & 10.7  & $  34.3 \pm  4.0$  &  $ 0.08^{+ 0.35}_{- 0.07}$  &  9395 &  17.6  &  40.70  & 1 \\
 J125946.28+312726.5  & 194.94283  &  31.45736  & 0.0508  & 97.3  & 10.5  & $  15.6 \pm  3.5$  &  $ 0.01^{+ 0.05}_{- 0.01}$  &  9395 &  17.6  &  40.69  & 0 \\
 J141503.63+520434.3  & 213.76512  &  52.07619  & 0.0732  & 98.8  & 10.8  & $  17.4 \pm  3.1$  &  $ 0.54^{+ 0.71}_{- 0.29}$  &  5854* & 195.8  &  40.16  & 1 \\
 J141503.92+520323.5  & 213.76633  &  52.05653  & 0.0730  & 98.8  & 10.7  & $  12.6 \pm  2.3$  &  $ 1.57^{+ 3.97}_{- 1.24}$  &  5854* & 195.8  &  40.16  & 1 \\
\enddata
\tablecomments{* represent the targets with multiple {\it Chandra} observations. The observation ID of all available observations are listed here (the ID listed in table is the one observation with the longest exposure time). \\
J1259+2757 555 556 1086 1112 1113 1114 9714 10672 13993 13994 13995 13996 14406 14410 14411 14415\\
J0320+4123 11715 11716 12037\\
J0924+0225 5604 11562 11563 11564 11565 11566\\
J1605+1746 4996 19592 20086 20087\\
J0805+2410 9270 19496 20888 20889 20890 20891\\
J1415+5204 5853 5854 6222 6223 6366 7187\\
(1) SDSS names with J2000 coordinates given in the form of "hhmmss.ss+ddmmss.s"; (2)-(3) Centroid position of the AGNs; (4) Redshift; (5) Projected physical separation of AGN in each pair, in units of kpc; (6) Stellar mass, in units of $M_{\odot}$; (7) [O\,III] emission line fluxes and 1-$\sigma$ uncertainties, in units of $10^{-17}{\rm~erg~s^{-1}~cm^{-2}}$; (8) Star formation rate, in units of $\rm{M}_{\odot}{\rm~{yr}^{-1}}$, given by the MPA-JHU DR7 catalog inferred from D$_n$(4000); (9) {\it Chandra} observation ID; (10) {\it Chandra} effective exposure time, in units of ks; (11) 0.5-8 keV limiting luminosity for source detection, in units of ${\rm~erg~s^{-1}}$; (12) Flag for X-ray detection, 1 and 0 represent detection and non-detection in X-ray, respectively. 
}
\end{deluxetable}

\startlongtable
\begin{deluxetable}{ccccccccc}
\tabletypesize{\scriptsize}
\tablecaption{X-ray properties of AGN pairs  \label{tab:Xray}}
\tablewidth{0pt}
\tablehead{
\colhead{Name} & \colhead{XR.A.} & \colhead{XDec.} & \colhead{Counts} & \colhead{$ \rm F_{0.5-2}$} & \colhead{$\rm F_{2-8}$} &  \colhead{log $L_{\rm 0.5-2}$} & \colhead{log $L_{\rm 2-10}$} & 
\colhead{HR}
}
\colnumbers
\startdata
  J145059.71-000215.2  & 222.74890  &  -0.03757&  $  57.6^{+ 8.1}_{- 8.0}$  &  $  7.62^{+ 1.23}_{- 1.12}$  &  $  1.31^{+ 0.50}_{- 0.41}$  &  $40.86^{+0.07}_{-0.07}$  &  $40.68^{+0.14}_{-0.16}$  &  $-0.65^{+0.09}_{-0.12}$ \\
  J145059.84-000213.2  & 222.74933  &  -0.03700&  $   5.5^{+ 3.1}_{- 2.5}$  &  $<$   2.20  &  $<$  1.45  &  $<$ 40.32  &  $<$ 40.73     &  $-0.19^{+0.34}_{-0.57}$ \\
  J102700.56+174900.3  & 156.75225  &  17.81680&  $  46.9^{+ 8.1}_{- 7.4}$  &  $  2.18^{+ 0.40}_{- 0.37}$  &  $  0.70^{+ 0.24}_{- 0.21}$  &  $40.71^{+0.07}_{-0.08}$  &  $40.81^{+0.13}_{-0.15}$  &  $-0.55^{+0.11}_{-0.14}$ \\
  J085837.53+182221.6  & 134.65693  &  18.37263&  $  15.7^{+ 4.7}_{- 4.0}$  &  $  2.12^{+ 0.62}_{- 0.53}$  &  $<$  0.97  &  $40.58^{+0.11}_{-0.13}$  &  $<$ 40.83  &  $-0.96^{+0.00}_{-0.04}$ \\
  J085837.68+182223.4  & 134.65700  &  18.37317&  $  13.6^{+ 4.4}_{- 3.7}$  &  $  1.69^{+ 0.56}_{- 0.47}$  &  $<$  1.31  &  $40.49^{+0.12}_{-0.14}$  &  $<$ 40.97  &  $-0.84^{+0.04}_{-0.16}$ \\
  J154403.45+044607.5  & 236.01437  &   4.76875&  $   5.0^{+ 2.9}_{- 2.2}$  &  $  1.17^{+ 0.55}_{- 0.43}$  &  $<$  1.72  &  $40.02^{+0.17}_{-0.20}$  &  $<$ 40.78  &  $-0.74^{+0.07}_{-0.26}$ \\
  J154403.67+044610.1  & 236.01530  &   4.76934&  $  59.3^{+ 8.2}_{- 8.1}$  &  $  1.33^{+ 0.59}_{- 0.47}$  &  $ 10.19^{+ 1.51}_{- 1.49}$  &  $40.07^{+0.16}_{-0.19}$  & $41.55^{+0.06}_{-0.07}$  &  $ 0.73^{+0.11}_{-0.07}$ \\
  J105842.44+314457.6  & 164.67689  &  31.74925&  $   2.6^{+ 2.3}_{- 1.7}$  &  $  0.36^{+ 0.36}_{- 0.24}$  &  $<$  1.74  &  $40.01^{+0.30}_{-0.46}$  &  $<$ 41.28  &  $-0.42^{+0.19}_{-0.58}$ \\
  J105842.58+314459.8  & 164.67743  &  31.74989&  $  94.0^{+10.3}_{-10.2}$  &  $  1.18^{+ 0.56}_{- 0.44}$  &  $ 18.20^{+ 2.04}_{- 2.02}$  &  $40.52^{+0.17}_{-0.20}$  &  $42.29^{+0.05}_{-0.05}$  &  $ 0.85^{+0.07}_{-0.04}$ \\
  J002208.69+002200.5  &   5.53621  &   0.36681&  $  16.7^{+ 5.1}_{- 4.4}$  &  $  0.75^{+ 0.23}_{- 0.20}$  &  $<$  0.68  &  $40.30^{+0.12}_{-0.13}$  &  $<$ 40.85  &  $-0.77^{+0.08}_{-0.23}$ \\
  J002208.83+002202.8  &   5.53679  &   0.36744&  $  15.5^{+ 4.9}_{- 4.3}$  &  $  0.65^{+ 0.22}_{- 0.19}$  &  $<$  0.78  &  $40.24^{+0.13}_{-0.15}$  &  $<$ 40.91  &  $-0.65^{+0.12}_{-0.26}$ \\
  J220635.08+000323.2  & 331.64613  &   0.05650&  $  76.9^{+ 9.3}_{- 9.2}$  &  $  1.57^{+ 0.73}_{- 0.59}$  &  $ 13.25^{+ 1.68}_{- 1.66}$  &  $40.24^{+0.17}_{-0.21}$  &  $41.75^{+0.05}_{-0.06}$  &  $ 0.81^{+0.09}_{-0.05}$ \\
  J133032.00-003613.5  & 202.63348  &  -0.60376&  $  16.2^{+ 4.7}_{- 4.0}$  &  $  2.09^{+ 0.71}_{- 0.59}$  &  $  1.55^{+ 0.64}_{- 0.54}$  &  $40.50^{+0.13}_{-0.15}$  &  $40.96^{+0.15}_{-0.19}$  &  $-0.20^{+0.21}_{-0.24}$ \\
  J121345.95+024839.0  & 183.44148  &   2.81086&  $  11.3^{+ 4.0}_{- 3.4}$  &  $<$   3.19  &  $  1.91^{+ 0.88}_{- 0.71}$  &  $<$ 40.96  &  $41.33^{+0.16}_{-0.20}$  &  $ 0.19^{+0.32}_{-0.32}$ \\
  J121346.11+024841.4  & 183.44183  &   2.81139&  $  14.9^{+ 4.5}_{- 3.9}$  &  $  1.90^{+ 0.76}_{- 0.62}$  &  $  1.59^{+ 0.81}_{- 0.64}$  &  $40.73^{+0.15}_{-0.17}$  &  $41.25^{+0.18}_{-0.22}$  &  $-0.23^{+0.23}_{-0.29}$ \\
  J141447.15-000013.3  & 213.69647  &  -0.00376&  $1198.5^{+36.6}_{-36.2}$  &  $ 27.32^{+ 1.44}_{- 1.43}$  &  $ 40.60^{+ 1.52}_{- 1.50}$  &  $41.50^{+0.02}_{-0.02}$  &  $42.27^{+0.02}_{-0.02}$  &  $ 0.33^{+0.03}_{-0.03}$ \\
  J141447.48-000011.3  & 213.69784  &  -0.00320&  $ 562.2^{+25.1}_{-24.9}$  &  $ 22.69^{+ 1.32}_{- 1.31}$  &  $ 11.83^{+ 0.82}_{- 0.82}$  &  $41.42^{+0.02}_{-0.03}$  &  $41.73^{+0.03}_{-0.03}$  &  $-0.18^{+0.04}_{-0.04}$ \\
  J080523.29+281815.8  & 121.34708  &  28.30438&  $  30.8^{+ 6.2}_{- 5.6}$  &  $  3.68^{+ 0.95}_{- 0.82}$  &  $  2.60^{+ 0.85}_{- 0.73}$  &  $41.54^{+0.10}_{-0.11}$  &  $41.97^{+0.12}_{-0.14}$  &  $-0.23^{+0.17}_{-0.19}$ \\
  J133817.27+481632.3  & 204.57208  &  48.27566&  $  95.2^{+10.8}_{-10.6}$  &  $ 16.18^{+ 2.08}_{- 2.06}$  &  $  4.99^{+ 1.47}_{- 1.26}$  &  $40.80^{+0.05}_{-0.06}$  &  $40.88^{+0.11}_{-0.13}$  &  $-0.65^{+0.07}_{-0.09}$ \\
  J133817.77+481641.1  & 204.57415  &  48.27808&  $ 205.0^{+15.3}_{-15.2}$  &  $ 25.62^{+ 2.53}_{- 2.51}$  &  $ 27.09^{+ 3.11}_{- 3.08}$  &  $41.00^{+0.04}_{-0.04}$  &  $41.61^{+0.05}_{-0.05}$  &  $-0.17^{+0.06}_{-0.08}$ \\
  J114753.63+094552.0  & 176.97338  &   9.76443&  $3343.2^{+61.0}_{-60.4}$  &  $102.63^{+ 4.11}_{- 4.07}$  &  $353.07^{+ 7.24}_{- 7.17}$  &  $42.70^{+0.02}_{-0.02}$  &  $43.83^{+0.01}_{-0.01}$  &  $ 0.58^{+0.01}_{-0.01}$ \\
  J005113.94+002047.2  &  12.80803  &   0.34644&  $  33.2^{+ 6.4}_{- 5.8}$  &  $  2.57^{+ 0.59}_{- 0.58}$  &  $  1.25^{+ 0.42}_{- 0.36}$  &  $41.26^{+0.09}_{-0.11}$  &  $41.53^{+0.13}_{-0.15}$  &  $-0.27^{+0.16}_{-0.19}$ \\
  J005114.12+002049.2  &  12.80872  &   0.34709&  $   6.4^{+ 3.1}_{- 2.4}$  &  $  0.39^{+ 0.28}_{- 0.20}$  &  $  0.34^{+ 0.24}_{- 0.17}$  &  $40.43^{+0.24}_{-0.32}$  &  $40.98^{+0.23}_{-0.31}$  &  $ 0.02^{+0.39}_{-0.40}$ \\
  J085312.36+162619.5  & 133.30145  &  16.43874&  $  16.5^{+ 4.6}_{- 4.0}$  &  $  3.72^{+ 1.08}_{- 0.94}$  &  $<$  2.55  &  $40.90^{+0.11}_{-0.13}$  &  $<$ 41.33  &  $-0.82^{+0.04}_{-0.17}$ \\
  J085312.70+162615.5  & 133.30362  &  16.43762&  $  92.8^{+10.2}_{-10.1}$  &  $  2.84^{+ 0.97}_{- 0.82}$  &  $ 24.14^{+ 2.83}_{- 2.80}$  &  $40.80^{+0.13}_{-0.15}$  &  $42.32^{+0.05}_{-0.05}$  &  $ 0.74^{+0.08}_{-0.06}$ \\
  J083848.15+040734.1  & 129.70055  &   4.12617&  $ 717.9^{+28.3}_{-28.0}$  &  $ 10.48^{+ 1.66}_{- 1.52}$  &  $136.41^{+ 5.57}_{- 5.51}$  &  $41.09^{+0.06}_{-0.07}$  &  $42.79^{+0.02}_{-0.02}$  &  $ 0.86^{+0.02}_{-0.02}$ \\
  J083817.59+305453.5  & 129.57327  &  30.91486&  $   7.8^{+ 2.9}_{- 2.8}$  &  $  5.23^{+ 2.16}_{- 2.03}$  &  $<$  5.88  &  $40.79^{+0.15}_{-0.21}$  &  $<$ 41.43  &  $-0.64^{+0.09}_{-0.35}$ \\
  J090714.45+520343.4  & 136.81035  &  52.06204&  $  41.9^{+ 7.2}_{- 6.5}$  &  $  0.99^{+ 0.52}_{- 0.40}$  &  $  7.51^{+ 1.30}_{- 1.29}$  &  $40.27^{+0.18}_{-0.22}$  &  $41.74^{+0.07}_{-0.08}$  &  $ 0.72^{+0.13}_{-0.08}$ \\
  J090714.61+520350.7  & 136.81095  &  52.06411&  $ 120.9^{+11.6}_{-11.5}$  &  $  4.82^{+ 1.02}_{- 0.95}$  &  $ 19.25^{+ 2.09}_{- 2.07}$  &  $40.96^{+0.08}_{-0.10}$  &  $42.15^{+0.04}_{-0.05}$  &  $ 0.55^{+0.09}_{-0.07}$ \\
  J134736.41+173404.7  & 206.90177  &  17.56804&  $  67.8^{+ 8.7}_{- 8.6}$  &  $ 84.17^{+11.27}_{-11.15}$  &  $  9.67^{+ 4.47}_{- 4.05}$  &  $41.94^{+0.05}_{-0.06}$  &  $41.59^{+0.17}_{-0.24}$  &  $-0.82^{+0.05}_{-0.08}$ \\
  J000249.07+004504.8  &   0.70424  &   0.75133&  $  13.0^{+ 4.2}_{- 3.6}$  &  $  2.86^{+ 1.77}_{- 1.33}$  &  $  6.22^{+ 2.49}_{- 2.04}$  &  $41.07^{+0.21}_{-0.27}$  &  $41.99^{+0.15}_{-0.17}$  &  $ 0.33^{+0.31}_{-0.23}$ \\
 J135429.06+132757.3  & 208.62109  &  13.46601&  $ 233.1^{+16.1}_{-16.0}$  &  $  2.74^{+ 1.10}_{- 0.90}$  &  $ 73.81^{+ 5.20}_{- 5.15}$  &  $40.76^{+0.15}_{-0.17}$  &  $42.78^{+0.03}_{-0.03}$  &  $ 0.92^{+0.03}_{-0.02}$ \\
 J124610.11+304354.9  & 191.54192  &  30.73191&  $  18.3^{+ 5.4}_{- 4.7}$  &  $  1.56^{+ 0.44}_{- 0.39}$  &  $  0.66^{+ 0.38}_{- 0.29}$  &  $39.57^{+0.11}_{-0.12}$  &  $39.79^{+0.20}_{-0.25}$  &  $-0.55^{+0.15}_{-0.23}$ \\
 J124611.24+304321.9  & 191.54684  &  30.72281&  $  70.6^{+ 9.0}_{- 9.0}$  &  $  3.27^{+ 0.54}_{- 0.49}$  &  $  2.45^{+ 0.56}_{- 0.51}$  &  $39.89^{+0.07}_{-0.07}$  &  $40.36^{+0.09}_{-0.10}$  &  $-0.34^{+0.10}_{-0.13}$ \\
 J111519.98+542316.7  & 168.83312  &  54.38791&  $1496.8^{+40.9}_{-40.4}$  &  $ 17.10^{+ 1.61}_{- 1.60}$  &  $214.35^{+ 6.11}_{- 6.05}$  &  $41.65^{+0.04}_{-0.04}$  &  $43.34^{+0.01}_{-0.01}$  &  $ 0.83^{+0.01}_{-0.02}$ \\
 J111830.28+402554.0  & 169.62611  &  40.43169&  $1859.6^{+45.5}_{-45.1}$  &  $288.97^{+ 8.03}_{- 7.95}$  &  $ 88.81^{+ 4.58}_{- 4.54}$  &  $43.60^{+0.01}_{-0.01}$  &  $43.68^{+0.02}_{-0.02}$  &  $-0.55^{+0.02}_{-0.02}$ \\
 J214623.23+000456.7  & 326.59679  &   0.08242&  $   8.3^{+ 3.7}_{- 3.0}$  &  $<$   2.74  &  $  2.13^{+ 0.97}_{- 0.79}$  &  $<$ 40.92  &  $41.40^{+0.16}_{-0.20}$  &  $ 0.80^{+0.20}_{-0.04}$ \\
 J085441.74+305754.7  & 133.67393  &  30.96521&  $ 556.5^{+24.9}_{-24.7}$  &  $155.97^{+ 8.10}_{- 8.02}$  &  $ 52.30^{+ 4.63}_{- 4.58}$  &  $43.56^{+0.02}_{-0.02}$  &  $43.68^{+0.04}_{-0.04}$  &  $-0.49^{+0.04}_{-0.04}$ \\
 J142947.66+353427.5  & 217.44857  &  35.57429&  $ 251.6^{+16.8}_{-16.6}$  &  $ 18.99^{+ 2.39}_{- 2.36}$  &  $ 48.03^{+ 3.79}_{- 3.75}$  &  $40.89^{+0.05}_{-0.06}$  &  $41.88^{+0.03}_{-0.04}$  &  $ 0.43^{+0.06}_{-0.06}$ \\
 J155549.45+242911.5  & 238.95602  &  24.48637&  $  35.5^{+ 6.5}_{- 6.1}$  &  $  1.57^{+ 0.58}_{- 0.48}$  &  $  4.44^{+ 0.94}_{- 0.87}$  &  $40.62^{+0.14}_{-0.16}$  &  $41.66^{+0.08}_{-0.10}$  &  $ 0.44^{+0.17}_{-0.13}$ \\
 J125929.96+275723.2  & 194.87483  &  27.95644&  $ 154.2^{+31.3}_{-31.5}$  &  $  0.97^{+ 0.20}_{- 0.20}$  &  $  0.39^{+ 0.14}_{- 0.13}$  &  $39.40^{+0.08}_{-0.10}$  &  $39.59^{+0.13}_{-0.18}$  &  $-0.48^{+0.25}_{-0.20}$ \\
 J125934.12+275648.6  & 194.89207  &  27.94666&  $ 334.7^{+30.5}_{-30.2}$  &  $  2.03^{+ 0.18}_{- 0.18}$  &  $<$  0.49  &  $39.77^{+0.04}_{-0.04}$  &  $<$ 39.74  &  $-0.82^{+0.10}_{-0.08}$ \\
 J131513.87+442426.5  & 198.80783  &  44.40742&  $  14.5^{+ 4.5}_{- 3.9}$  &  $  1.03^{+ 0.34}_{- 0.28}$  &  $<$  1.05  &  $39.82^{+0.12}_{-0.14}$  &  $<$ 40.42  &  $-0.73^{+0.07}_{-0.24}$ \\
 J131517.27+442425.6  & 198.82213  &  44.40720&  $3835.5^{+65.4}_{-64.8}$  &  $ 17.35^{+ 1.26}_{- 1.24}$  &  $363.46^{+ 6.37}_{- 6.31}$  &  $41.05^{+0.03}_{-0.03}$  &  $42.96^{+0.01}_{-0.01}$  &  $ 0.89^{+0.01}_{-0.01}$ \\
 J142627.54+330412.8  & 216.61463  &  33.07008&  $   2.1^{+ 2.0}_{- 1.3}$  &  $<$   2.28  &  $<$  1.53  &  $<$ 40.70  &  $<$ 41.12     &  $-0.04^{+0.56}_{-0.61}$ \\
 J011544.85+001400.0  &  18.93686  &   0.23343&  $ 367.9^{+20.4}_{-20.2}$  &  $ 29.79^{+ 2.05}_{- 2.03}$  &  $ 16.68^{+ 1.55}_{- 1.54}$  &  $41.48^{+0.03}_{-0.03}$  &  $41.82^{+0.04}_{-0.04}$  &  $-0.28^{+0.05}_{-0.05}$ \\
 J112545.05+144035.7  & 171.43785  &  14.67653&  $  25.2^{+ 5.7}_{- 5.0}$  &  $  4.16^{+ 1.01}_{- 0.89}$  &  $  0.68^{+ 0.47}_{- 0.34}$  &  $40.39^{+0.09}_{-0.10}$  &  $40.19^{+0.23}_{-0.31}$  &  $-0.71^{+0.09}_{-0.18}$ \\
 J112549.55+144006.6  & 171.45646  &  14.66846&  $  17.4^{+ 4.8}_{- 4.2}$  &  $  3.32^{+ 0.91}_{- 0.79}$  &  $<$  1.34  &  $40.29^{+0.10}_{-0.12}$  &  $<$ 40.48  &  $-0.96^{+0.00}_{-0.04}$ \\
 J143310.55+525830.5  & 218.29404  &  52.97510&  $  13.4^{+ 3.6}_{- 3.9}$  &  $  5.63^{+ 3.01}_{- 2.56}$  &  $ 14.23^{+ 4.94}_{- 4.87}$  &  $40.82^{+0.19}_{-0.26}$  &  $41.81^{+0.13}_{-0.18}$  &  $ 0.31^{+0.29}_{-0.22}$ \\
 J143312.96+525747.3  & 218.30400  &  52.96314&  $   3.3^{+ 2.3}_{- 1.7}$  &  $  2.83^{+ 2.37}_{- 1.71}$  &  $<$ 10.45  &  $40.52^{+0.26}_{-0.40}$  &  $<$ 41.67  &  $-0.71^{+0.03}_{-0.29}$ \\
 J032015.52+412355.1  &  50.06477  &  41.39862&  $ 266.2^{+30.1}_{-29.8}$  &  $  5.00^{+ 0.53}_{- 0.52}$  &  $<$  1.28  &  $40.10^{+0.04}_{-0.05}$  &  $<$ 40.10  &  $-0.87^{+0.05}_{-0.06}$ \\
 J032512.84+003711.6  &  51.30350  &   0.61989&  $   <  10.8 $  &  $  1.83^{+ 1.52}_{- 1.11}$  &  $<$  7.19  &  $41.38^{+0.26}_{-0.40}$  &  $<$ 42.56  &  $-0.71^{+0.03}_{-0.29}$ \\
 J135740.18+223137.7  & 209.41768  &  22.52686&  $   8.6^{+ 3.5}_{- 2.8}$  &  $  1.19^{+ 0.88}_{- 0.62}$  &  $  2.06^{+ 0.97}_{- 0.77}$  &  $40.39^{+0.24}_{-0.32}$  &  $41.21^{+0.17}_{-0.20}$  &  $ 0.32^{+0.35}_{-0.26}$ \\
 J152131.56+074425.8  & 230.38150  &   7.74050&  $  40.4^{+ 9.1}_{- 8.4}$  &  $  2.06^{+ 0.51}_{- 0.46}$  &  $<$  1.72  &  $40.28^{+0.10}_{-0.11}$  &  $<$ 40.79  &  $-0.60^{+0.19}_{-0.19}$ \\
 J160003.55+412845.4  & 240.01480  &  41.47940&  $   8.7^{+ 3.5}_{- 2.9}$  &  $  1.25^{+ 0.66}_{- 0.50}$  &  $  1.62^{+ 0.86}_{- 0.66}$  &  $39.85^{+0.18}_{-0.22}$  &  $40.56^{+0.18}_{-0.23}$  &  $-0.01^{+0.31}_{-0.31}$ \\
 J092458.83+022519.7  & 141.24512  &   2.42214&  $  26.6^{+ 7.5}_{- 6.9}$  &  $  0.71^{+ 0.18}_{- 0.16}$  &  $<$  0.74  &  $40.29^{+0.10}_{-0.11}$  &  $<$ 40.89  &  $-0.75^{+0.09}_{-0.25}$ \\
 J150457.12+260058.5  & 226.23804  &  26.01631&  $  68.0^{+ 9.5}_{- 9.4}$  &  $  2.65^{+ 0.38}_{- 0.38}$  &  $  0.40^{+ 0.18}_{- 0.15}$  &  $40.61^{+0.06}_{-0.07}$  &  $40.37^{+0.16}_{-0.20}$  &  $-0.78^{+0.07}_{-0.10}$ \\
 J150501.22+260101.5  & 226.25512  &  26.01699&  $  15.1^{+ 4.6}_{- 4.0}$  &  $  0.39^{+ 0.17}_{- 0.14}$  &  $  0.24^{+ 0.15}_{- 0.11}$  &  $39.78^{+0.15}_{-0.19}$  &  $40.15^{+0.21}_{-0.27}$  &  $-0.32^{+0.23}_{-0.32}$ \\
 J123407.84+062922.5  & 188.53273  &   6.48972&  $   6.4^{+ 3.1}_{- 2.4}$  &  $  2.95^{+ 1.55}_{- 1.20}$  &  $<$  3.44  &  $41.02^{+0.18}_{-0.23}$  &  $<$ 41.67  &  $-0.69^{+0.09}_{-0.31}$ \\
 J151022.87+074304.2  & 227.59545  &   7.71802&  $   4.3^{+ 2.6}_{- 1.9}$  &  $  3.15^{+ 1.69}_{- 1.43}$  &  $<$  4.82  &  $40.50^{+0.19}_{-0.26}$  &  $<$ 41.28  &  $-0.84^{+0.00}_{-0.16}$ \\
 J211706.99-064327.9  & 319.27920  &  -6.72440&  $ 112.0^{+11.2}_{-11.1}$  &  $ 20.79^{+ 2.55}_{- 2.52}$  &  $  8.84^{+ 1.50}_{- 1.50}$  &  $40.95^{+0.05}_{-0.06}$  &  $41.17^{+0.07}_{-0.08}$  &  $-0.33^{+0.09}_{-0.10}$ \\
 J140343.63+292045.3  & 210.93175  &  29.34581&  $   4.3^{+ 2.6}_{- 2.0}$  &  $<$   7.20  &  $  3.73^{+ 2.38}_{- 1.92}$  &  $<$ 41.18  &  $41.49^{+0.21}_{-0.31}$  &  $ 0.45^{+0.50}_{-0.16}$ \\
 J140345.02+292144.0  & 210.93768  &  29.36215&  $  14.4^{+ 4.1}_{- 3.8}$  &  $  6.45^{+ 2.68}_{- 2.25}$  &  $  8.75^{+ 3.32}_{- 3.16}$  &  $41.14^{+0.15}_{-0.19}$  &  $41.86^{+0.14}_{-0.19}$  &  $ 0.00^{+0.26}_{-0.26}$ \\
 J121354.87+024753.0  & 183.47858  &   2.79806&  $  12.7^{+ 4.2}_{- 3.6}$  &  $  1.79^{+ 0.75}_{- 0.61}$  &  $  1.30^{+ 0.77}_{- 0.58}$  &  $40.72^{+0.15}_{-0.18}$  &  $41.17^{+0.20}_{-0.25}$  &  $-0.29^{+0.23}_{-0.31}$ \\
 J032227.88+421626.4  &  50.61617  &  42.27400&  $   4.3^{+ 2.6}_{- 1.9}$  &  $  3.23^{+ 2.07}_{- 1.66}$  &  $<$  6.16  &  $40.65^{+0.21}_{-0.31}$  &  $<$ 41.52  &  $-0.48^{+0.16}_{-0.42}$ \\
 J160501.37+174632.5  & 241.25571  &  17.77569&  $  < 32.4 $  &  $  0.93^{+ 0.39}_{- 0.34}$  &  $<$  0.63  &  $39.71^{+0.15}_{-0.20}$  &  $<$ 40.13  &  $-0.75^{+0.07}_{-0.25}$ \\
 J093325.71+340253.1  & 143.35742  &  34.04807&  $   8.4^{+ 4.7}_{- 4.0}$  &  $  1.11^{+ 0.50}_{- 0.43}$  &  $<$  0.63  &  $39.62^{+0.16}_{-0.21}$  &  $<$ 39.96  &  $-0.92^{+0.00}_{-0.08}$ \\
 J104232.05+050241.9  & 160.63361  &   5.04496&  $ 606.7^{+26.0}_{-25.8}$  &  $  0.66^{+ 0.22}_{- 0.19}$  &  $ 36.20^{+ 1.57}_{- 1.55}$  &  $39.39^{+0.13}_{-0.14}$  &  $41.72^{+0.02}_{-0.02}$  &  $ 0.96^{+0.01}_{-0.01}$ \\
 J091339.47+295934.7  & 138.41444  &  29.99291&  $  56.2^{+ 8.2}_{- 8.1}$  &  $  6.32^{+ 0.95}_{- 0.94}$  &  $<$  1.80  &  $40.21^{+0.06}_{-0.07}$  &  $<$ 40.25  &  $-0.92^{+0.02}_{-0.07}$ \\
 J215945.57-002712.0  & 329.93987  &  -0.45333&  $   7.6^{+ 3.5}_{- 2.9}$  &  $  3.18^{+ 1.77}_{- 1.35}$  &  $<$  7.33  &  $41.46^{+0.19}_{-0.24}$  &  $<$ 42.42  &  $-0.39^{+0.32}_{-0.41}$ \\
 J215947.31-002644.3  & 329.94696  &  -0.44544&  $  21.7^{+ 5.3}_{- 4.7}$  &  $  1.51^{+ 1.06}_{- 0.76}$  &  $  8.33^{+ 2.23}_{- 1.94}$  &  $41.15^{+0.23}_{-0.31}$  &  $42.48^{+0.10}_{-0.11}$  &  $ 0.67^{+0.20}_{-0.11}$ \\
 J082321.67+042220.9  & 125.84048  &   4.37237&  $  50.1^{+ 8.4}_{- 8.3}$  &  $ 23.03^{+ 4.09}_{- 3.74}$  &  $<$  4.09  &  $41.05^{+0.07}_{-0.08}$  &  $<$ 40.89  &  $-0.93^{+0.02}_{-0.07}$ \\
 J080529.88+241004.4  & 121.37450  &  24.16789&  $  10.5^{+ 4.2}_{- 3.6}$  &  $<$   0.23  &  $  0.16^{+ 0.07}_{- 0.05}$  &  $<$ 39.63  &  $40.07^{+0.15}_{-0.18}$  &  $ 0.64^{+0.27}_{-0.11}$ \\
 J080535.00+240950.3  & 121.39592  &  24.16397&  $ 540.2^{+24.7}_{-24.4}$  &  $  2.71^{+ 0.25}_{- 0.24}$  &  $  6.83^{+ 0.36}_{- 0.36}$  &  $40.71^{+0.04}_{-0.04}$  &  $41.70^{+0.02}_{-0.02}$  &  $ 0.48^{+0.04}_{-0.04}$ \\
 J135419.95+325547.8  & 208.58317  &  32.92994&  $1244.1^{+37.2}_{-36.8}$  &  $263.45^{+ 9.21}_{- 9.12}$  &  $138.58^{+ 7.99}_{- 7.91}$  &  $41.96^{+0.01}_{-0.02}$  &  $42.27^{+0.02}_{-0.03}$  &  $-0.46^{+0.03}_{-0.02}$ \\
 J125939.94+312707.4  & 194.91642  &  31.45206&  $   7.3^{+ 3.5}_{- 2.9}$  &  $  0.88^{+ 0.58}_{- 0.43}$  &  $<$  2.74  &  $40.08^{+0.22}_{-0.29}$  &  $<$ 41.17  &  $-0.21^{+0.47}_{-0.41}$ \\
 J141503.63+520434.3  & 213.76512  &  52.07619&  $  24.4^{+ 6.2}_{- 5.6}$  &  $  0.30^{+ 0.10}_{- 0.08}$  &  $<$  0.41  &  $39.94^{+0.12}_{-0.14}$  &  $<$ 40.65  &  $-0.34^{+0.24}_{-0.25}$ \\
 J141503.92+520323.5  & 213.76633  &  52.05653&  $  26.9^{+ 6.9}_{- 6.2}$  &  $  0.41^{+ 0.11}_{- 0.10}$  &  $<$  0.45  &  $40.07^{+0.11}_{-0.12}$  &  $<$ 40.70  &  $ -0.47^{+0.27}_{-0.23}$ \\
\enddata
\tablecomments{(1) SDSS names with J2000 coordinates given in the form of "hhmmss.ss+ddmmss.s"; (2)-(3) Centroid position of the X-ray counterpart; (4) Observed net counts in 0.5-8 ($F$) keV bands; (5)-(6) Observed photon flux in 0.5-2 ($S$) and 2-8 ($H$) keV bands, in units of $10^{-6}{\rm~ph~cm^{-2}~s^{-1}}$; (7)-(8) 0.5-2 and 2--10 keV unabsorbed luminosities, in units of ${\rm erg~s^{-1}}$; (9) Hardness ratio between the 0.5-2 and 2-8 keV bands. 
}
\end{deluxetable}

\begin{deluxetable}{ccccccc}
\tabletypesize{\scriptsize}
\tablecaption{X-ray detection rate of different samples \label{tab:rate}}
\tablewidth{0pt}
\tablehead{
\colhead{Sample} & \colhead{Sample size} & \colhead{\# of detection} & \colhead{$F$ band} & \colhead{$S$ band} & \colhead{$H$ band} & \colhead{detection rate}
}
\colnumbers
\startdata
AGN pairs & 134 &  78 &   76 & 72 & 49 & $58\% \pm 7\%$  \\
Single AGNs & 115 &  66 &    60 & 58 & 36 & $57\% \pm 7\%$  \\
SFG pairs & 134 &  23 &    22 & 20 & 6 & $17\% \pm 4\%$  \\
SFG pairs (stellar mass $> 10^{9} \rm~M_{\sun}$)& 78 &  16 &   16  & 14 & 5 & $21\% \pm 5\%$  \\
AGN pairs ($r_p \lesssim 10$ kpc) & 40 &  31 &   31  & 29 &  22& $78\% \pm 14\%$  \\
AGN pairs ($r_p \gtrsim 10$ kpc) & 94 &  47 &   45  & 43 & 27 & $50\% \pm 7\%$  \\
AGN pairs ($L_{\rm 2-10} > 10^{41} \rm~erg~s^{-1}$) & 134 &  36 &   36  & 33 & 36 & $27\% \pm 4\%$  \\
Single AGNs ($L_{\rm 2-10} > 10^{41} \rm~erg~s^{-1}$) & 115 & 28  &   28  & 24 & 28 & $24\% \pm 5\%$  \\
SFG pairs ($L_{\rm 2-10} > 10^{41} \rm~erg~s^{-1}$) & 134 & 2  &   2  & 1 & 2 & $1\% \pm 1\%$  \\
\hline
(detection limit $< 10^{41.2} \rm~erg~s^{-1}$) \\
AGN pairs & 114 &  68 &    67 & 63 & 42 & $60\% \pm 7\%$  \\
Single AGNs & 94 &  59 &    54 & 53 & 32 & $63\% \pm 8\%$  \\
SFG pairs & 90 &  20 &    20 & 17 & 6 & $22\% \pm 5\%$  \\
\enddata
\tablecomments{(1) Samples; (2) Number of nuclei in different samples;  (3) Number of detected nuclei in X-ray for different samples; (4)-(6) Number of detected nuclei in the $F$, $S$ and $H$ X-ray band, respectively; (7) X-ray detection rate.
}
\end{deluxetable}

\begin{deluxetable}{lccccccc}
\tablecaption{X-ray Luminosity of the three newly confirmed kpc-scale AGN pairs \label{table:newlylx}}
\tabletypesize{\footnotesize}
\tablehead{
\colhead{Name} & 
\colhead{$r_{\rm p}$} &
\colhead{Counts} & 
\colhead{log${L_{0.5-2}}$} & 
\colhead{log${L_{2-10}}$} & 
\colhead{SFR} & 
\colhead{log${L^{\rm SF}_{\rm 0.5-2}}$} & 
\colhead{log${L^{\rm SF}_{\rm 2-10}}$} 
}
\colnumbers
\startdata
 J1450-0002a & 2.3&  $  57.6^{+ 8.1}_{- 8.0}$  &  $40.86^{+0.07}_{-0.07}$  &  $40.68^{+0.14}_{-0.16}$  &  $ 0.11^{+ 0.45}_{- 0.11}$  &  $38.71^{+0.69}_{-1.37}$  &  $38.76^{+0.69}_{-1.37  }$  \\
 J1450-0002b & 2.3&  $   5.5^{+ 3.1}_{- 2.5}$  &  $<$ 40.32 &  $<$ 40.73   &  $ 0.02^{+ 0.10}_{- 0.02}$  &  $37.85^{+0.87}_{-1.31}$  &  $37.90^{+0.87}_{-1.31  }$  \\
 J0858+1822a & 3.3&  $  15.7^{+ 4.7}_{- 4.0}$  &  $40.58^{+0.11}_{-0.13}$  &  $<$ 40.83  &  $ 3.55^{+ 8.31}_{- 2.62}$  &  $40.20^{+0.52}_{-0.58}$  &  $40.25^{+0.52}_{-0.58  }$  \\
 J0858+1822b & 3.3&  $  13.6^{+ 4.4}_{- 3.7}$  &  $40.49^{+0.12}_{-0.14}$  &  $<$ 40.97  &  $ 3.46^{+ 9.05}_{- 2.88}$  &  $40.19^{+0.56}_{-0.78}$  &  $40.24^{+0.56}_{-0.78  }$  \\
 J1414-0000a & 4.9&  $1198.6^{+36.6}_{-36.2}$  &  $41.50^{+0.02}_{-0.02}$  &  $42.27^{+0.02}_{-0.02}$  &  $ 0.27^{+ 0.77}_{- 0.24}$  &  $39.08^{+0.59}_{-1.03}$  &  $39.12^{+0.59}_{-1.03  }$  \\
 J1414-0000b & 4.9&  $ 562.2^{+25.1}_{-24.9}$  &  $41.42^{+0.02}_{-0.03}$  & $41.73^{+0.03}_{-0.03}$  &  $ 3.54^{+ 4.94}_{- 2.12}$  &  $40.20^{+0.38}_{-0.40}$  &  $40.25^{+0.38}_{-0.40  }$  \\
\enddata
\tablecomments{(2) Projected physical separation of AGN in each pair, in units of kpc; (3) Observed net counts in 0.5-8 ($F$) keV bands; (4)-(5) 0.5-2 ($S$) and 2--10 keV bands unabsorbed luminosities, in units of ${\rm erg~s^{-1}}$; (6) Fiber star formation rate, in units of $\rm{M}_{\odot}{\rm~{yr}^{-1}}$, given by the MPA-JHU DR7 catalog inferred from D$_n$(4000); (7)-(8) 0.5-2 ($S$) and 2--10 keV bands X-ray luminosities due to star formation. 
}
\end{deluxetable}

\end{document}